\documentclass[prx,twocolumn,amsmath,amssymb,dvipdfmx]{revtex4}

\usepackage{color}
\usepackage{bm}
\usepackage{dcolumn}
\usepackage[dvipdfmx]{graphicx}
\usepackage{slashed}
\usepackage{amsmath}\usepackage{accents}
\usepackage{ulem}

\newcommand{\cin}[1]{{#1}}
\newcommand{\cinb}[1]{{#1}}

\newcommand{\1}{\mbox{1}\hspace{-0.25em}\mbox{l}}

\newcommand{\tr}{{\rm tr}\,}

\begin{document}

\title{
Boundary-obstructed topological phases of a \cin{massive} Dirac fermion in a magnetic field
}

\author{Koichi Asaga and Takahiro Fukui}
\affiliation{Department of Physics, Ibaraki University, Mito 310-8512, Japan}

\date{\today}

\begin{abstract}
It is known that in some higher-order topological insulators \cin{(HOTIs)}, \cin{topological} phases are distinguished not by 
gap closings of bulk states but by those of edge states,
which are called boundary-obstructed topological phases (BOTPs).
In this paper, we 
\cinb{
construct an effective theory of
} 
the BOTP 
\cinb{
transition
} 
of  two-dimensional (2D) Su-Schrieffer-Heeger (SSH)  model in a uniform magnetic field.
At $\pi$ flux per plaquette, 
this model corresponds to the typical model of \cin{HOTIs} proposed by Benalcazar, Bernevig, and Hughes (BBH).
The BBH model can be approximated by Dirac fermions with two kinds of mass terms, 
which will be referred to as BBH Dirac \cin{insulator}.
To clarify the BOTP 
\cinb{
transition
}
of the 2D SSH model around $\pi$ flux, 
we study \cin{such BBH Dirac insulator in the presence of} a magnetic field.
\cin{On the other hand, generically}
in continuum Dirac models, boundary conditions associated with the Hermiticity of Hamiltonians
are known to play a crucial role in determining the edge states. 
We first demonstrate that for the  conventional Dirac fermion with a single mass term, 
such boundary conditions indeed determine the edge states even in the presence of a magnetic field.
Next, imposing boundary conditions consistent to \cin{the lattice terminations and symmetries of the BBH Hamiltonian
as well as to the Hermiticity of the BBH Dirac insulator}, 
we obtain the edge states of \cin{the BBH Dirac insulator} 
in a magnetic field and 
\cinb{
reproduce
} 
its BOTP 
\cinb{
transition.
}
In particular, we show that the unpaired Landau levels, which cause the spectral asymmetry, 
yield the edge states responsible for the BOTP 
\cinb{
transition.
}
\end{abstract}

\pacs{
}

\maketitle

\section{Introduction}

Higher-order topological insulators  \cin{(HOTIs)} 
\cite{Benalcazar:2017aa,Benalcazar:2017ab,Schindler:2018ab,Hayashi:2018aa,Hashimoto:2016aa,Hashimoto:2017aa}
have been attracting much current interest 
\cite{Langbehn:2017aa,Song:2017aa,Ezawa:2018aa,Ezawa:2018ab,Liu:2017aa,
Khalaf:2018cr,Matsugatani:2018aa,Fukui:2018aa,Calugaru:2019aa}.
While conventional (first-order) topological insulators (TI) accompany bulk gap closings in their topological transitions,
\cinb{
some
} 
\cin{HOTIs} can change topological properties without bulk gap closings: 
Instead, gap closings of edge states induce \cin{topological} changes generically, implying that
\cinb{ 
those
} 
 \cin{HOTI} phases are distinguished by gap closings of edge states.
Such properties, called boundary-obstructed topological phases (BOTP), 
have been studied in Ref. \cite{1908.00011}.
\cinb{
On the other hand, the breathing kagome lattice model \cite{Ezawa:2018aa} is one example of HOTIs
with bulk gap closings.
}

One of typical examples 
\cinb{
showing the BOTP
} 
is the two-dimensional (2D) second-order 
topological quadrupole model proposed by Benalcazar, Bernevig, and Hughes (BBH) 
\cite{Benalcazar:2017aa,Benalcazar:2017ab}.
This model is a kind of 2D generalization of the one-dimensional (1D) Su-Schrieffer-Heeger (SSH) model \cite{Su:1979aa}.
The BBH model has been further generalized by introducing locally oscillating flux of zero mean \cite{Wheeler:2019tg} 
or uniform flux \cite{Otaki:2019aa},
both of which interpolate the 2D SSH model with zero flux  and 
with $\pi$ flux per plaquette.
It has been pointed out in Ref. \cite{Otaki:2019aa} that as a function of the magnetic flux (Hofstadter butterfly),
there appear many gapped regions at half-filling showing corner states.
In particular, around $\pi$ flux, relatively a  large gap is open, whose ground states are expected to \cin{be continuously 
connected with the ground state of the BBH model without any gap closings.}
Thus, if anisotropic hoppings breaking C$_4$ symmetry
are introduced, those ground states would reveal the BOTP.

In this paper, we investigate the BOTP \cite{1908.00011}
of the anisotropic BBH model \cite{Benalcazar:2017aa,Benalcazar:2017ab} in a magnetic field.
To this end, we use Dirac \cin{insulator description} in the continuum limit associated with 
high-symmetry points of the BBH model \cite{Benalcazar:2017aa,Fukui:2019aa}.
\cin{It has been pointed out that such continuum models are composed of doubled Dirac fermions incorporated by $4\times4$ $\gamma$ matrices
with two kinds of mass terms, which will be referred to as BBH Dirac insulator. Previously, the same model has been 
studied in the context of superconducting Dirac fermions in a vortex background \cite{JackiwRossi:1981}.
Thus, our motivation in this paper is to study the BOTP 
\cinb{
transition
}
of the BBH Dirac insulator {\it in a magnetic field}.}

\cin{When we discuss edge states of continuum fermions,
 it is known that the boundary condition 
ensuring the Hermiticity of their Hamiltonians plays a crucial role \cite{1510.07698,Hashimoto:2016aa,Hashimoto:2017aa}.
Therefore, apart from the present BBH model or BBH Dirac insulator model, we first investigate edge states of the 
conventional 2D Dirac fermion with a single mass term in a magnetic field, imposing a generic boundary condition
allowed by Hermiticity of the Hamiltonian.
We show that among Landau levels, the unpaired Landau level yields  an edge state 
approximately equivalent to the one in the absence of a magnetic field.
For the bulk system, this unpaired Landau level is known to yield the unit Hall conductivity  \cite{Ishikawa:1985uq}.
It is also related with the parity anomaly of the massive Dirac fermion in three dimensions 
\cite{alvarez85,Ishikawa:1984aa,Semenoff:1984aa,Redlich:1984kx,Redlich:1984uq}.
We show that even for a system with a boundary, the behavior of the edge state associated with the unpaired Landau level  
is of great importance: It is the only one edge state that can cross the zero energy.}

 \cin{Based on these results, we next proceed to study the BOTP 
 \cinb{
transition
}
 of the BBH Dirac insulator. 
Since the BBH Dirac insulator is derived from the BBH model on the lattice,}
we emphasize the importance of its boundary conditions
required 
(1) by the boundary termination \cin{of the BBH model} on the lattice,
(2) by symmetries of the BBH model,
and 
(3) by the  Hermiticity of the \cin{BBH Dirac insulator Hamiltonian.}
Based on exact and/or numerical solutions for edge states, we  
\cinb{ 
show that the BOTP transition occurs in the BBH Dirac insulator in a magnetic field.
}
\cin{Namely, due to the doubling of massive Dirac fermions in the BBH Dirac insulator model, 
chiral edge states of massive Dirac fermions couple together, forming gapped edge states.
These become 1D topological insulators, whose mass gap closing induces the topological transition,
as already known for the BBH model in the absence of a magnetic field. 
We show  that in the presence of a magnetic 
field, the edge state of the unpaired Landau level plays the same role and causes the BOTP 
\cinb{
transition.
}
Although the other Landau levels also yield edge states due to their nontrivial Chern numbers, 
the unpaired Landau level is solely relevant to the BOTP \cin{transition}.  
} 

This paper is organized as follows:
The next Sec. \ref{s:BBH} is devoted to the overview of the lattice BBH model and its continuum limit.
First, we give a brief review of the BBH model in Sec. \ref{s:bbh_lat} to fix our notational conventions, 
and second, taking the continuum limit of the lattice model, 
we derive the BBH Dirac \cin{insulator model} in a magnetic field in Sec. \ref{s:continuum}, 
including discussions of the boundary conditions in Sec. \ref{s:bbh_bou}.  
As argued in 
\cite{1510.07698,Hashimoto:2016aa,Hashimoto:2017aa},
Hamiltonians of continuum fermions are not necessarily Hermitian if boundaries are introduced. 
Then, when we determine the edge states,
boundary conditions 
which make the Hamiltonians Hermitian play a crucial role. 
Generically, such boundary conditions
allow \cin{parameter dependence, as will be discussed in Sec. \ref{s:simple}.} 
However, given a lattice model,
lattice terminations would choose unique boundary conditions, \cin{which naturally keeps the Hamiltonian Hermitian.}
We argue several aspects of the boundary conditions of the \cin{lattice} BBH model and BBH Dirac \cin{insulator} model. 

\begin{figure}[htb]
\begin{center}
\begin{tabular}{c}
\includegraphics[width=.5\linewidth]{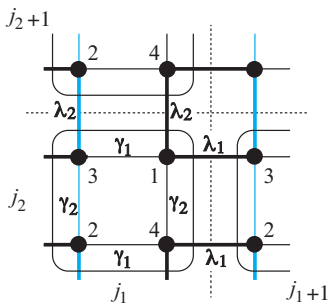}
\end{tabular}
\caption{
The BBH model ($\pi$ flux).  Blue lines are links to which  the phase $e^{i\pi}=-1$ denoting the $\pi$ flux are attached.
The dotted-lines are lattice terminations when we consider the boundaries.
}
\label{f:bbh}
\end{center}
\end{figure}

\begin{figure*}[htb]
\begin{center}
\begin{tabular}{ccccccc}
\includegraphics[width=.13\linewidth]{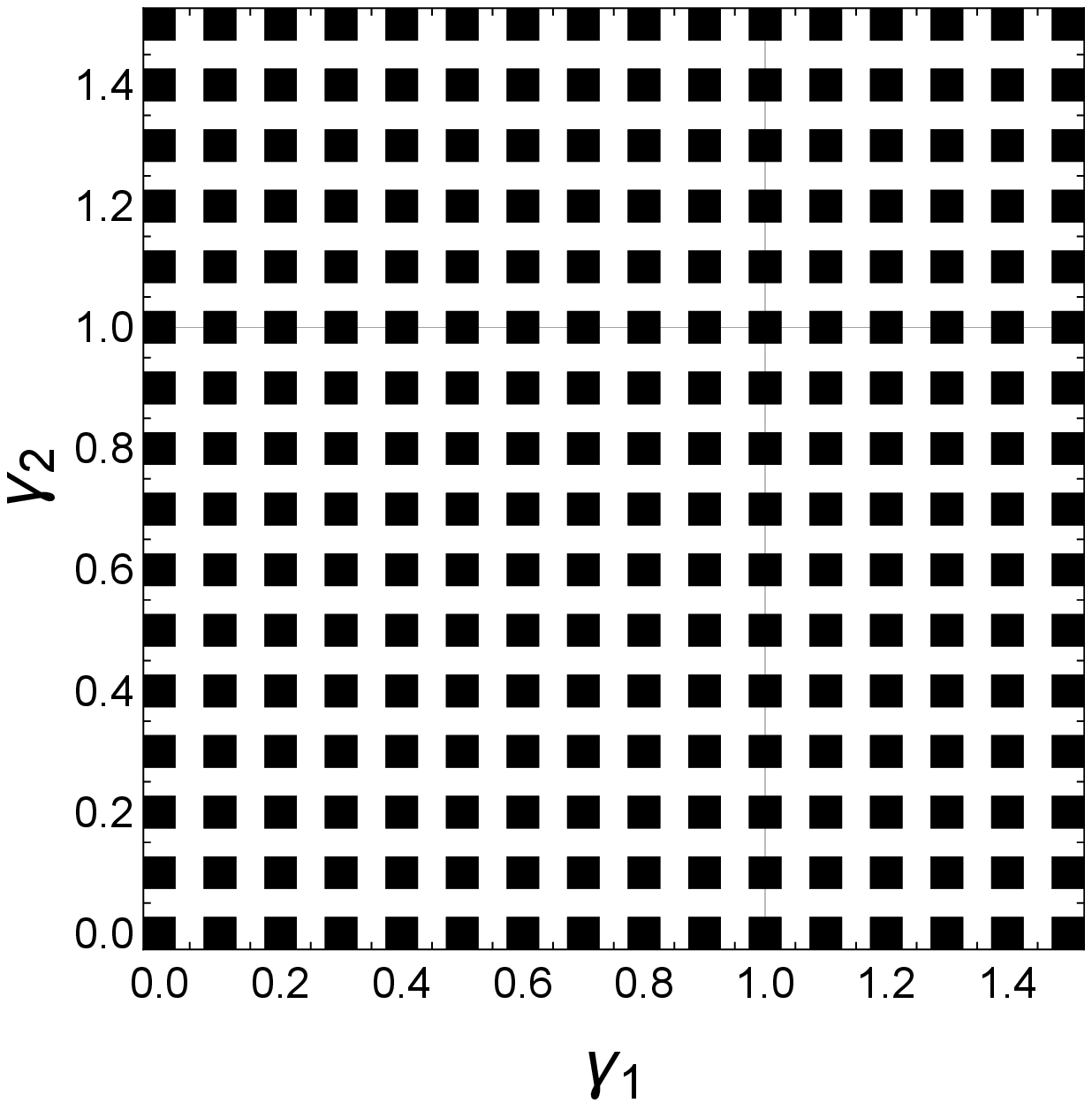}
&
\includegraphics[width=.13\linewidth]{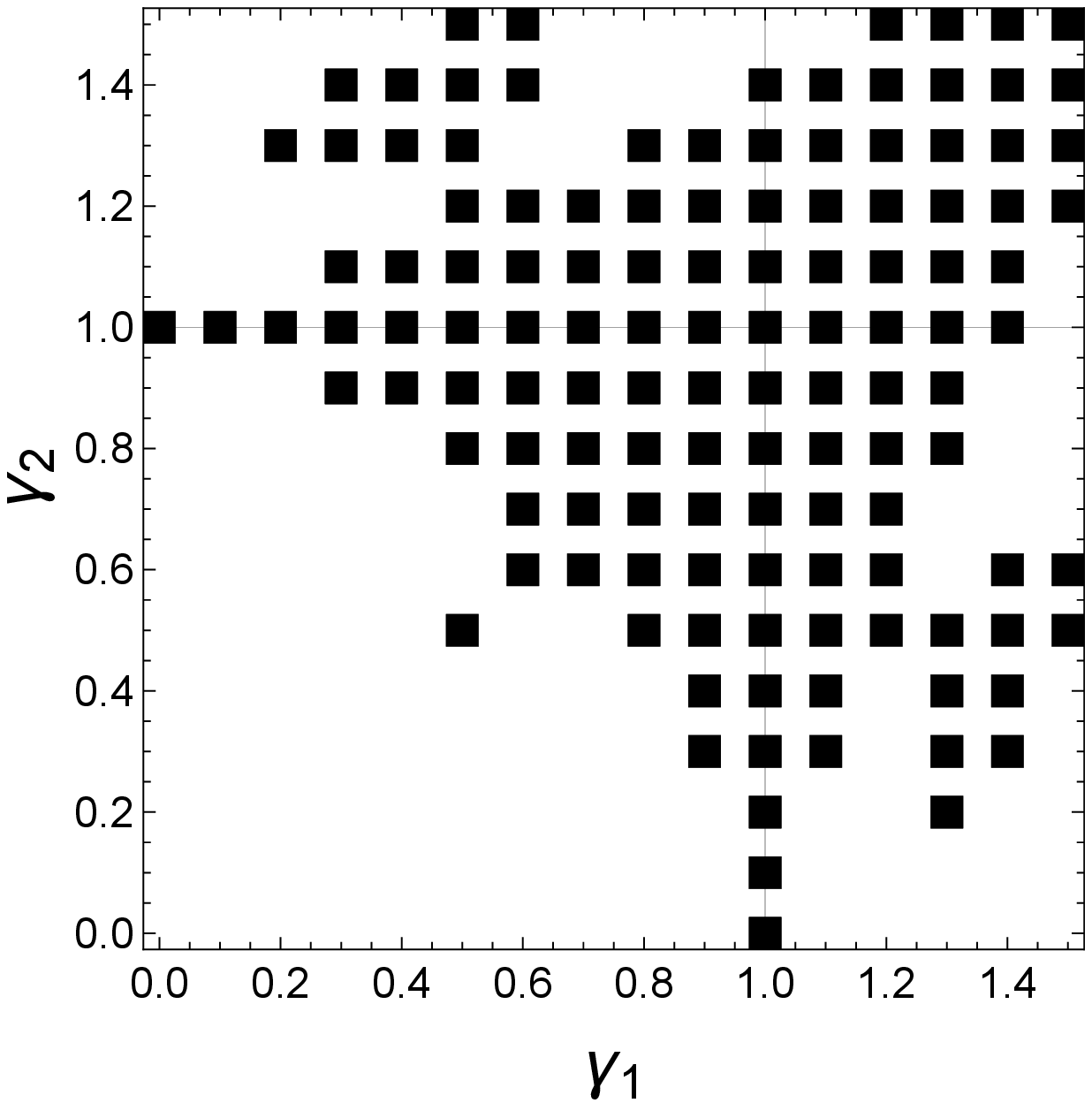}
&
\includegraphics[width=.13\linewidth]{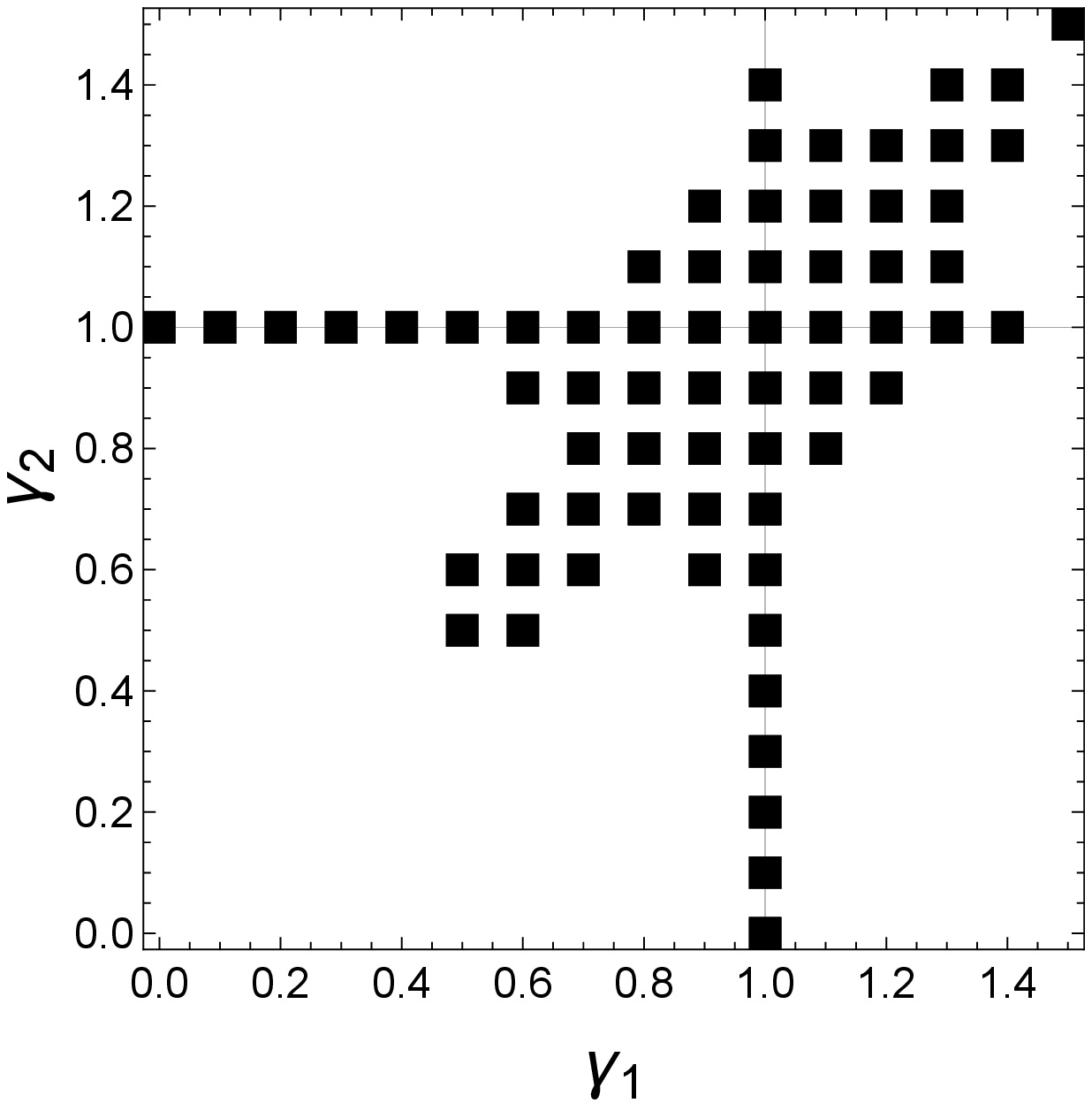}
&
\includegraphics[width=.13\linewidth]{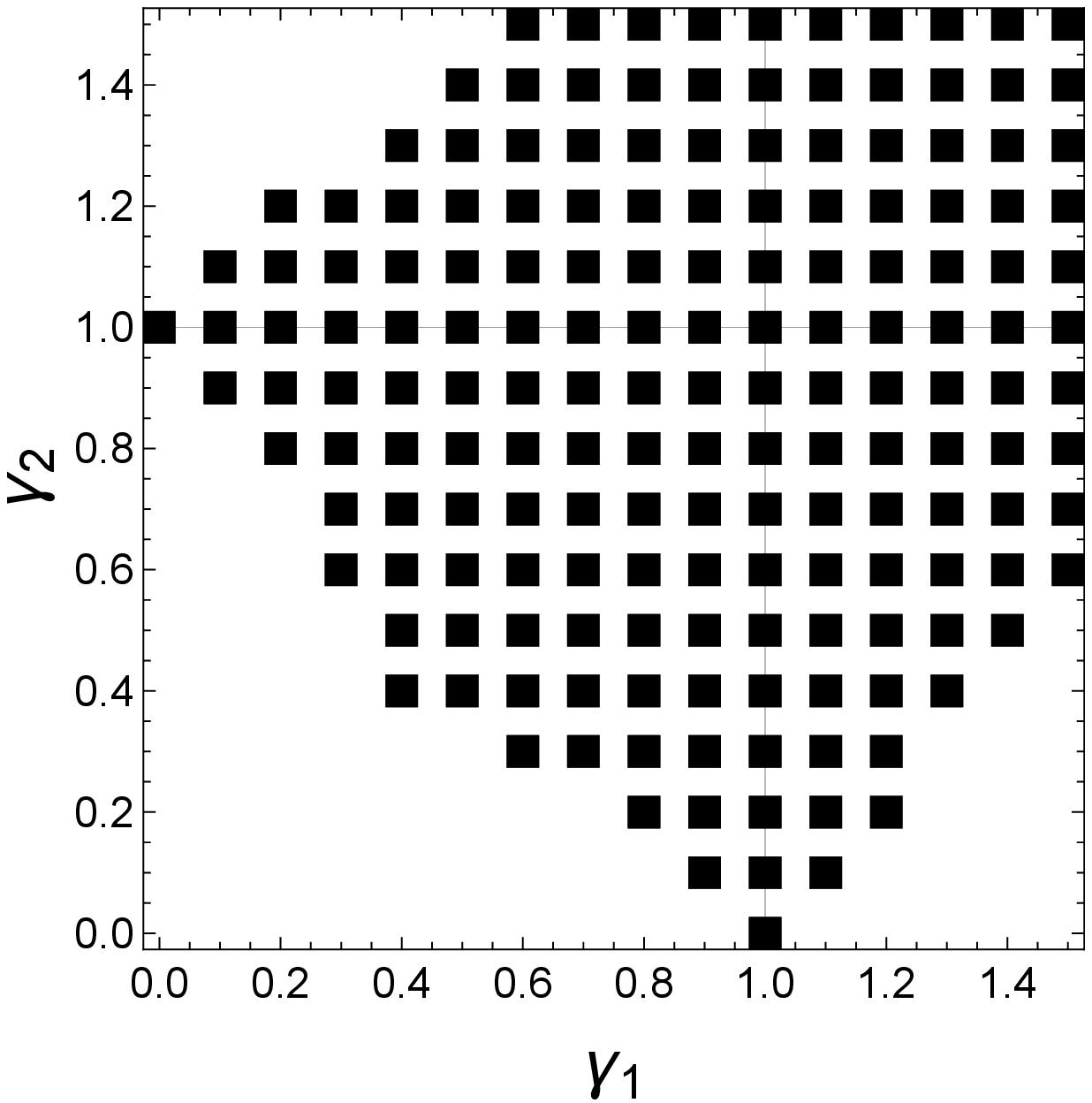}
&
\includegraphics[width=.13\linewidth]{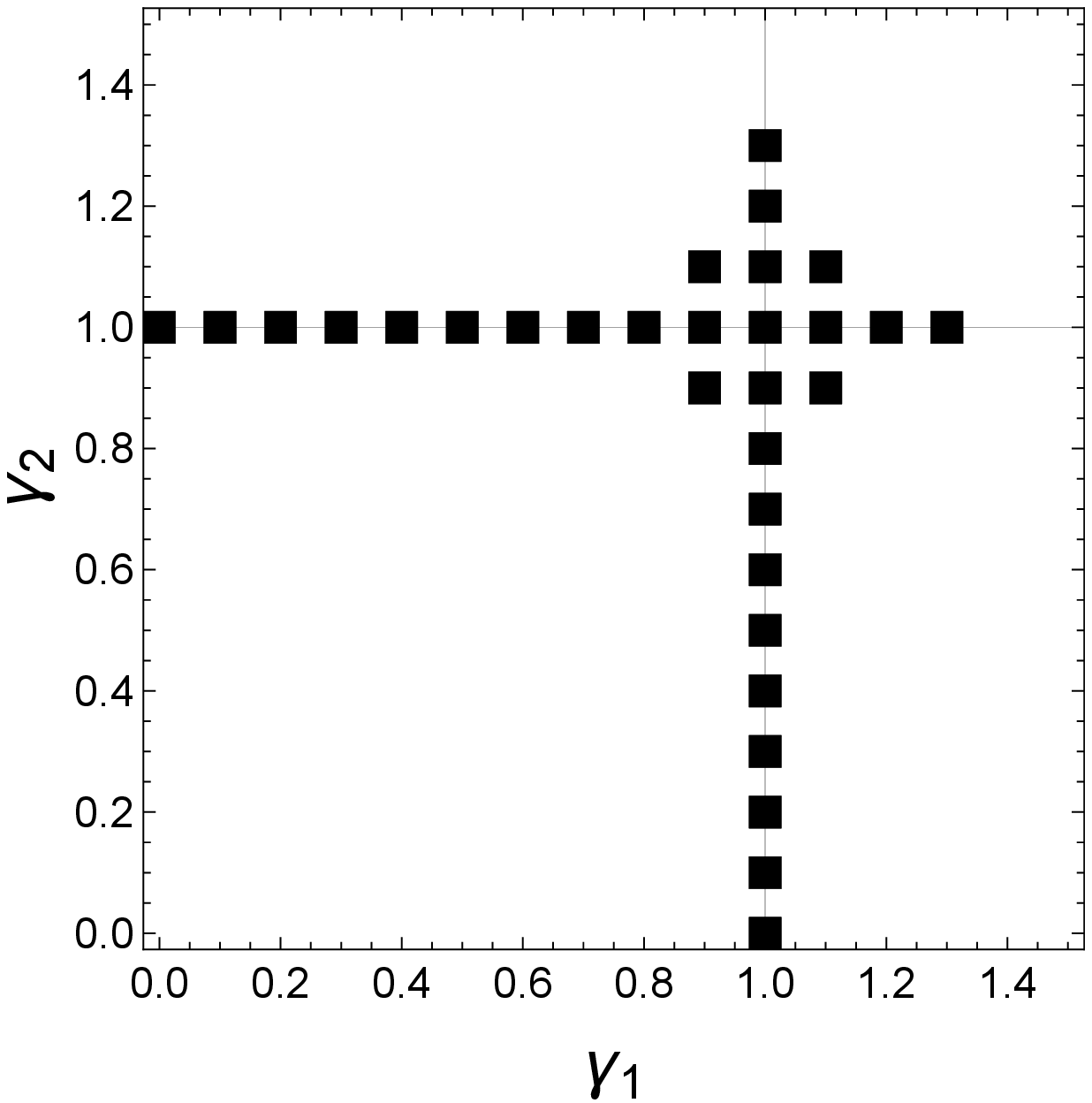}
&
\includegraphics[width=.13\linewidth]{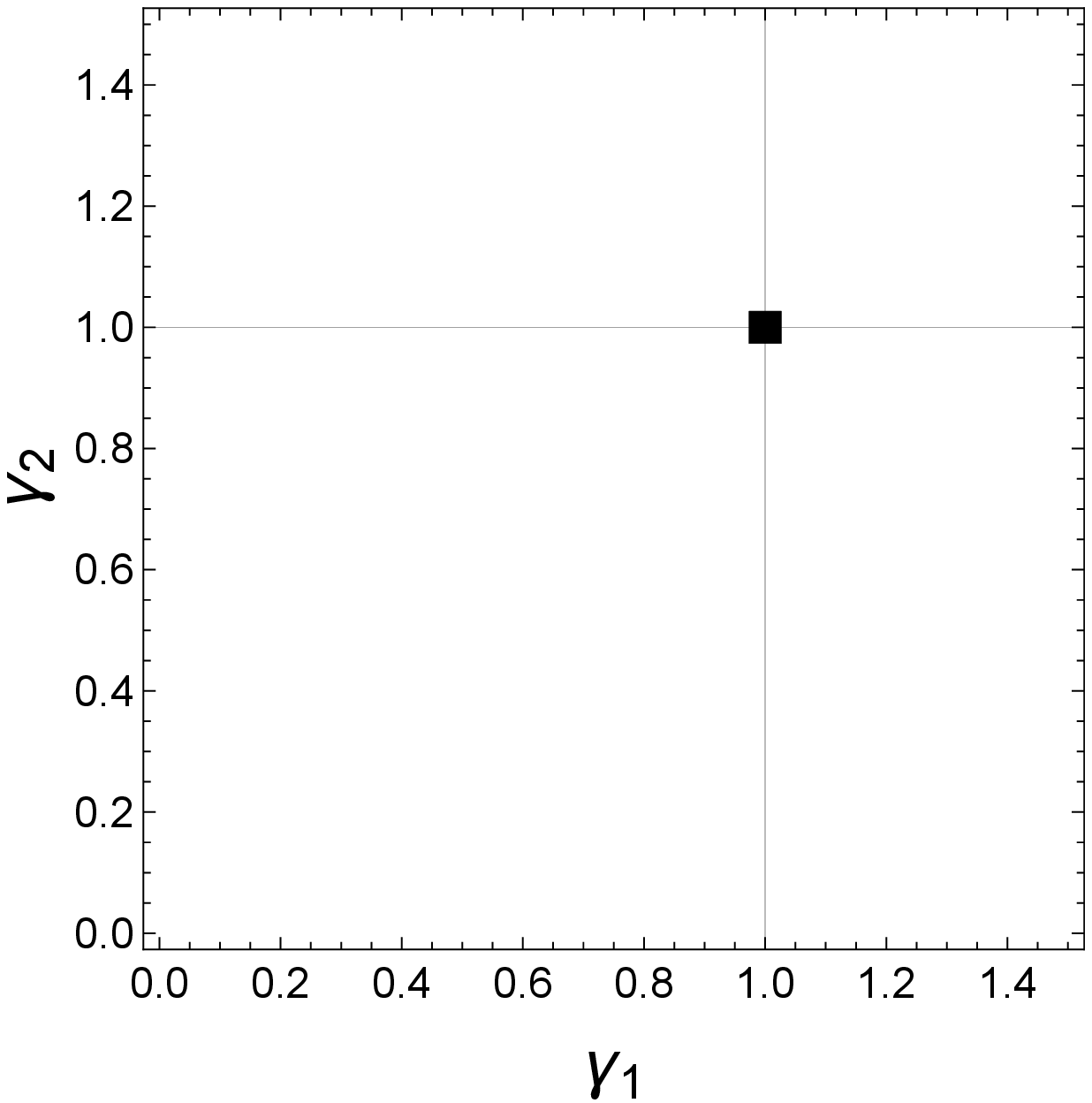}
&
\includegraphics[width=.13\linewidth]{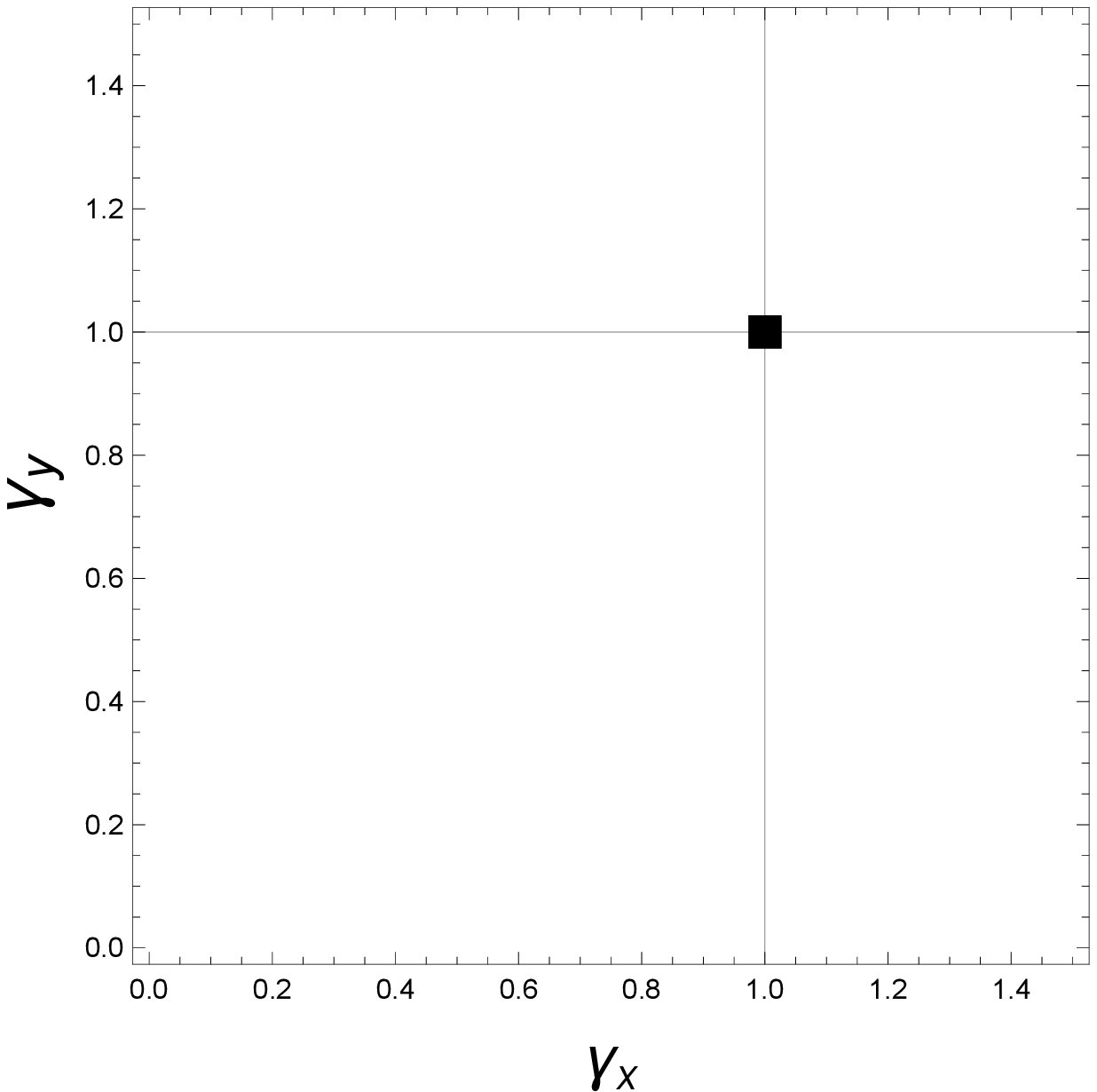}
\\
$\phi=0$
&$\phi=\frac{1}{6}\pi$
&$\phi=\frac{1}{3}\pi$&$\phi=\frac{1}{2}\pi$&$\phi=\frac{2}{3}\pi$&$\phi=\frac{5}{6}\pi$&$\phi=\pi$
\end{tabular}
\caption{
Bulk gapless points of the model with flux $\phi=\frac{p}{6}\pi$ with $p=0,1,\cdots,6$ shown on the $\gamma_1$-$\gamma_2$ 
($\lambda=1$) plane 
within the region
$0\leq\gamma_1,\gamma_2\leq1.5$.
The black squares indicate the gapless points, where the white regions show the gapful region.
The rightmost panel is the BBH model with $\pi$ flux. The second panel from the right is the model with flux $\phi=\frac{5}{6}\pi$, 
which can alternatively be regarded as the BBH model under a flux
$\phi'=\frac{5}{6}\pi-\pi=-\frac{1}{6}\pi$.
}
\label{f:BOTP}
\end{center}
\end{figure*}

Before discussing the BOTP of the 
BBH Dirac \cin{insulator}, we discuss the conventional 2D \cin{massive} Dirac fermion in Sec. \ref{s:simple}.
\cin{Without considering any concrete lattice models, 
we have to} take account of generic boundary conditions. In the former part, Sec. \ref{s:dir_nomag}, 
we give a brief review of edge states for the \cin{massive} Dirac fermion in the absence of a magnetic field, 
and in the latter part, Sec \ref{s:simple_mag}, \cin{we derive the edge states in the presence of a magnetic field.}
We show that among Landau levels of the \cin{massive} Dirac fermion, the unpaired level, which causes the spectral asymmetry,
is responsible for the edge states involved in topological properties of the model.

In Sec. \ref{s:bbh_dirac}, we switch to the BBH Dirac \cin{insulator model}.
We first argue, in Sec. \ref{s:BBHnomag}, the BOTP of the BBH Dirac \cin{insulator} in the absence of a magnetic field, 
although discussed already in Ref. \cite{Benalcazar:2017aa}, with particular emphasis on the boundary conditions.  
Next, in Sec. \ref{s:bbh_mag}, we discuss the BOTP 
\cinb{
transition
}
of the BBH Dirac \cin{insulator} in a magnetic field.
It is shown that the edge states associated with the unpaired Landau levels of the \cin{massive} Dirac fermion are responsible for
the BOTP of the BBH Dirac \cin{insulator}. 
Finally, in Sec. \ref{s:sum}, we give summary and discussion.

\section{BBH model}
\label{s:BBH}

In this section, we review basic properties of the BBH model in a uniform magnetic field.
The BBH model, which is originally a 2D SSH model with $\pi$ flux, has been generalized 
in Ref. \cite{Otaki:2019aa}, including arbitrary uniform flux. The model then interpolates 
a simple 2D SSH model with zero flux and the BBH model with $\pi$ flux.
It has been shown that around $\pi$ flux, there appear relatively large gap, whose ground states 
\cin{could be continuously deformed into the ground state of the BBH model without gap closings.}
To confirm this, we focus our attention on the \cin{HOTI} phase in such a large gap region around $\pi$ flux.

\subsection{Overview of the lattice model}\label{s:bbh_lat}

The BBH Hamiltonian on the lattice in Fig. \ref{f:bbh} is defined by
\begin{alignat}1
H&=\sum_j \Big[
\gamma_1(c_{1,j}^\dagger c_{3,j}+c_{2,j}^\dagger c_{4,j})+\lambda_1(c_{1,j}^\dagger c_{3,j+1}+c_{2,j+1}^\dagger c_{4,j})
\nonumber\\
&\qquad+\gamma_2(c_{1,j}^\dagger c_{4,j}-c_{2,j}^\dagger c_{3,j})+\lambda_2(c_{1,j}^\dagger c_{4,j+1}-c_{2,j+1}^\dagger c_{3,j})
\Big]
\nonumber\\
&\qquad+\mbox{H.c}
\nonumber\\
&=\sum_{i,j} c_i^\dagger {\cal H}^\ell_{ij} c_j,
\label{BBHHam2nd}
\end{alignat}
where $\gamma_j$ denotes the hopping within a unit cell, whereas $\lambda_j$ denotes 
the hopping between the unit cells in the $j=1,2$ direction, and simple $c_j=(c_{1,j},\cdots,c_{4,j})^T$ is the abbreviation of the multicomponent fermion annihilation operator.
The Fourier transformation 
leads to 
\begin{alignat}1
{\cal H}^\ell(k)&=\sum_{j=1}^4\Gamma^j g_j(k),
\label{BBHHam1}
\end{alignat}
where $g_j(k)$ 
is given by
$g_j(k)=\lambda_j\sin k_j$  $(j=1,2)$ and 
$g_{j+2}(k)=\gamma_j+\lambda_j \cos k_j$ $(j=1,2)$.
The $\Gamma$-matrices are defined by
$\Gamma^1=-\tau^2\sigma^3$, $\Gamma^2=-\tau^2\sigma^1$, $\Gamma^3=\tau^1\sigma^0$, 
and $\Gamma^4=-\tau^2\sigma^2$ as well as $\Gamma_5=-\tau^3\sigma^0$, where $\sigma^\mu$ and $\tau^\mu$
are conventional Pauli matrices with $\sigma^0=\tau^0=\1$.
They obey 
$\{\Gamma^j,\Gamma^l\}=2\delta^{jl}$ ($j,l=1,\cdots,4$) 
and $\Gamma_5=(-i)^2\Gamma^1\Gamma^2\Gamma^3\Gamma^4$, so that
$\tr\Gamma_5\Gamma^1\Gamma^2\Gamma^3\Gamma^4=(2i)^2 $.

For the BBH model, reflection symmetries play a crucial role:
\begin{alignat}1
&M_1{\cal H}^\ell(k_1,k_2)M_1^{-1}={\cal H}^\ell(-k_1,k_2),
\nonumber\\
&M_2{\cal H}^\ell(k_1,k_2)M_2^{-1}={\cal H}^\ell(k_1,-k_2),
\label{BBHSym}
\end{alignat}
where $M_1=i\Gamma^1\Gamma_5$, $M_2=i\Gamma^2\Gamma_5$.
\cin{These reflection symmetries ensure the quantization of the polarizations with respect to the $1$- and $2$-directions, 
$(p_1,p_2)$,
which serve as topological invariants characterizing the HOTI phase of the BBH model \cite{Benalcazar:2017aa,Benalcazar:2017ab}.}

Let us next consider the effect of a uniform magnetic field.
In Ref. \cite{Otaki:2019aa}, 
the 2D SSH model in a generic magnetic field has been studied. 
This study was restricted to the 
model with C$_4$ synmmetry, i.e., $\gamma_1=\gamma_2\equiv\gamma$ and $\lambda_1=\lambda_2$.
It has been shown that 
there appear several gapped regions in the Hofstadter butterfly whose half-filled ground states belong to
the second-order topological insulating
phase characterized by the nontrivial quantized entanglement polarizations. 
\cin{Let ${\cal H}^\ell(k_1,k_2,B)$ be the BBH Hamiltonian with a uniform magnetic field $B$. Then, 
the reflection symmetries as well as time reversal symmetry are denoted by
\begin{alignat}1
&M_1{\cal H}^\ell(k_1,k_2,B)M_1^{-1}={\cal H}^\ell(-k_1,k_2,-B),
\nonumber\\
&M_2{\cal H}^\ell(k_1,k_2,B)M_2^{-1}={\cal H}^\ell(k_1,-k_2,-B),
\nonumber\\
&T{\cal H}^\ell(k_1,k_2,B)T^{-1}={\cal H}^\ell(-k_1,-k_2,-B).
\label{BBHMagSym}
\end{alignat}
Then, it turned out that the transformation laws under $M_1T$ and $M_2T$
ensure that the entanglement polarizations, which are alternative topological invariants describing the HOTI \cite{Fukui:2018aa},
$(p_1^\sigma,p_2^\tau)$, are quantized even in the presence of a magnetic field \cite{Otaki:2019aa}, 
where $\sigma,\tau$ characterize the partitions of the unit cell. 
}

Now, let us relax the C$_4$ symmetry and compute the bulk energy gaps.
In Fig. \ref{f:BOTP}, we show gapless regions by black squares on the $\gamma_1$-$\gamma_2$ plane ($0\leq\gamma_j\leq1.5$),
where we have set $\lambda=1$.
Let $\phi$ be a magnetic flux per plaquette. 
Then, the rightmost panel ($\phi=\pi$) corresponds to the BBH model. Indeed, one can find that
the bulk gap closing occurs solely at $\gamma_1=\gamma_2=1$.
This is the BOTP 
\cinb{ 
transition:
} 
The ground state in the \cin{HOTI} phase ($\gamma_1,\gamma_2<1$) can be deformed into 
the trivial insulating phase ($\gamma_1>1$ and/or $\gamma_2>1$) without bulk gap closings.
Such a feature is not restricted to $\phi=\pi$:
The second panel from the right, which is the case with $\phi=\frac{5}{6}\pi$, is likewise,
suggesting that \cin{these ground states are topologically the same as the ground state of the BBH model. }

\cinb{
\begin{figure}[htb]
\begin{center}
\begin{tabular}{cc}
\includegraphics[width=.49\linewidth]{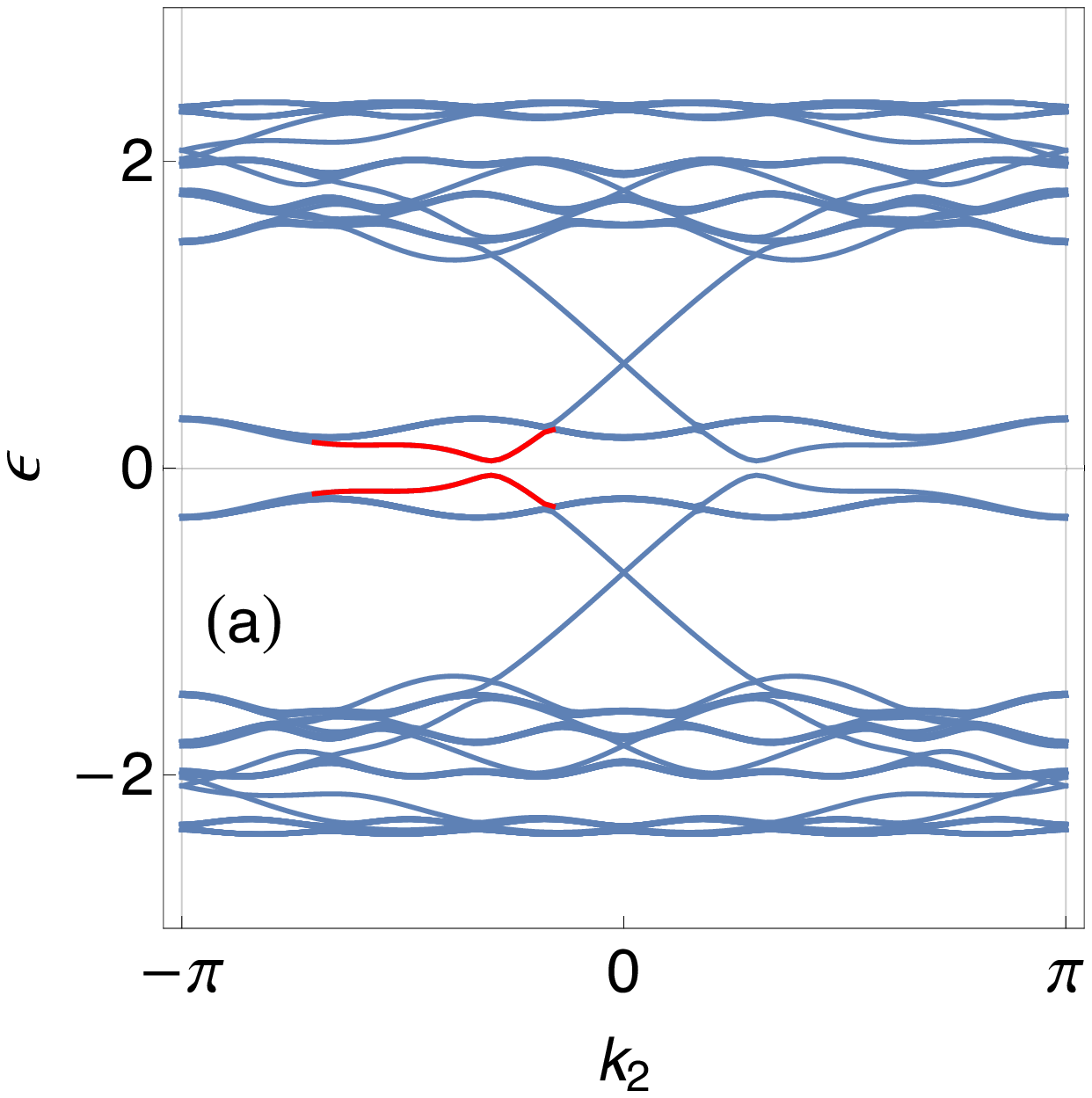}
&
\includegraphics[width=.49\linewidth]{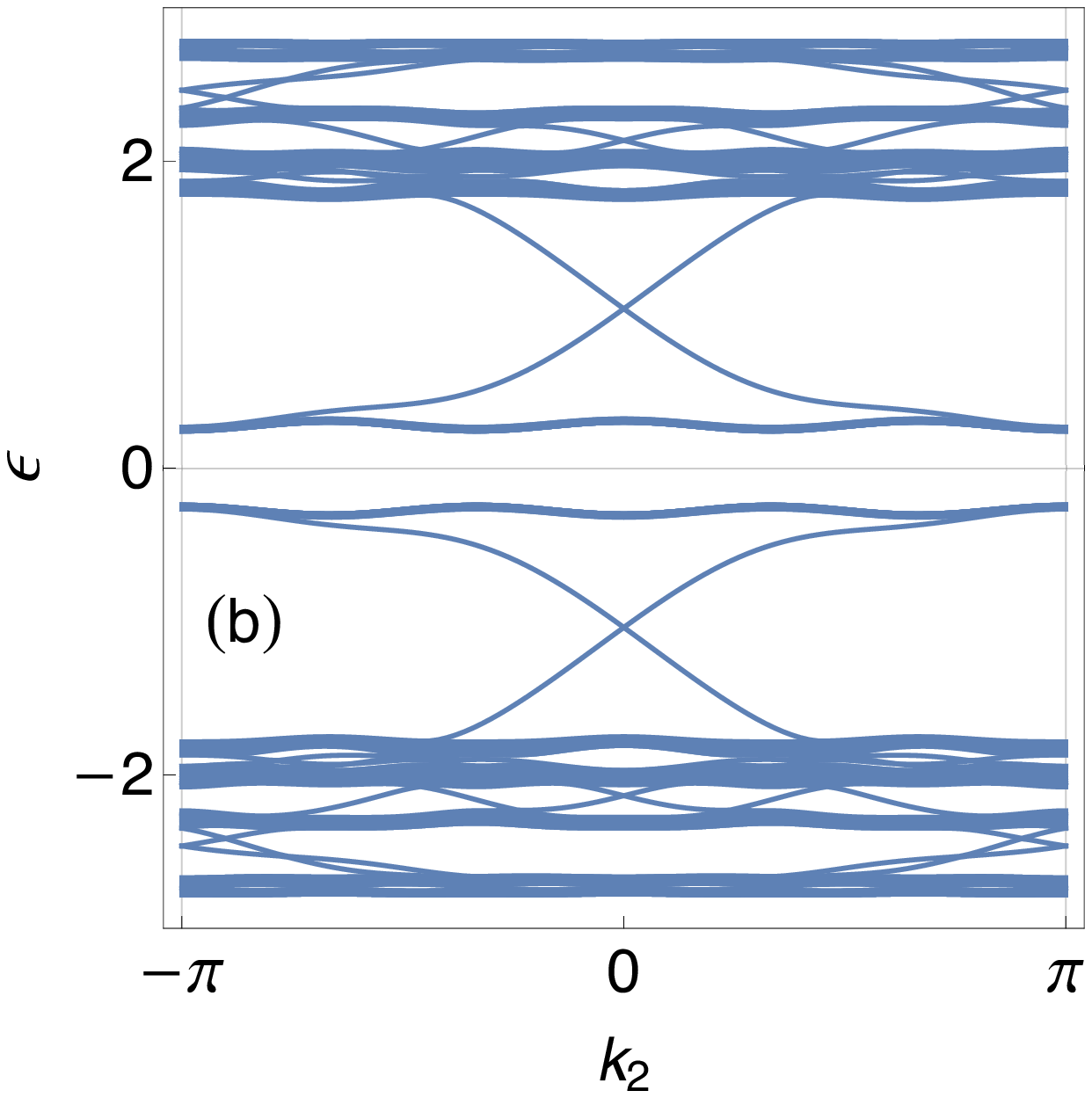}
\end{tabular}
\caption{
\cinb{ 
Spectra of the lattice model with $\phi=5\pi /6$. 
(a) $\gamma_1=0.7$ $\gamma_2=0.95$. The red curves are edge states localized at the left end. 
(b) $\gamma_1=1.3$ $\gamma_2=0.95$.
} 
}
\label{f:bbh_lat_edge}
\end{center}
\end{figure}
In Fig. \ref{f:bbh_lat_edge}, we show spectra of the lattice model in flux $\phi=5\pi/6$
with open boundary conditions in the 1-direction.
Figures \ref{f:bbh_lat_edge}(a) and (b) locate in the $\phi=5\pi/6$ panel in 
Fig. \ref{f:BOTP} at $\gamma_1<1$ and $\gamma_1>1$, respectively,
with the same $\gamma_2=0.95$ which is
just below the transition line of $\gamma_2=1$.
The half-filled ground states of these two cases can be distinguished by entanglement (bulk) polarizations 
proposed  in Refs. \cite{Fukui:2018aa,Otaki:2019aa},
$(p_1^\sigma,p_2^\tau)=(1/2,1/2)$ for (a), whereas (0,1/2) for (b), where
$(p_1^\sigma,p_2^\tau)$ in the magnetic unit cell
are similar to $(p_1^{(13)},p_2^{(14)})$  in Sec. \ref{s:ep} in the absence of a magnetic field.
It should be noted that 
the quantization of the entanglement polarizations is guaranteed by the symmetry properties (\ref{BBHMagSym})
\cite{Otaki:2019aa}.
The red curves in Fig. \ref{f:bbh_lat_edge}(a) stand for the edge states,
ensured by $p_1^\sigma=1/2$,
localized at the left end relevant to the BOTP transition.
These edge states are also characterized by  nontrivial entanglement (edge state) polarization
$1/2$.
If $\gamma_2$ passes $\gamma_2=1$, the gap between these states is closed and opens again
with the bulk gap kept open,
and  the entanglement edge state polarization changes into the trivial one.
This is the BOTP transition in the lattice model under a magnetic field.
On the other hand, 
in Fig. \ref{f:bbh_lat_edge}(b), we cannot observe any edge states within the bulk gap 
around zero energy due to $p_1^\sigma=0$.
Thus, the BOTP of the present system can be characterized by the entanglement  polarizations.

The purpose of this paper is to formulate an effective theory of the BOTP transition 
under a generic magnetic field around the BBH model.
To this end, we utilize Dirac \cin{insulator} models in the continuum limit of the BBH model and 
introduce a magnetic field to them. 
}

In passing, we mention that at $\phi=\frac{2}{3}\pi$, as shown in the third panel from the right, 
gapped ground states in $\gamma_1,\gamma_2<1$ region always accompany bulk gap closing across $\gamma_1=1$
or $\gamma_2=1$ lines.
From the point of view of symmetries, entanglement polarizations, and computed corner states, 
ground states with $\gamma_1,\gamma_2<1$ in this panel
belong to the \cin{HOTI} phase, but the \cin{topological} change is distinguished by bulk-gap closings 
like first-order TI.
Although this phase is out of the scope of this paper, it may be an interesting  issue to 
clarify the nature of this phase.

\subsection{Continuum limit}\label{s:continuum}

The lattice model (\ref{BBHHam1}) includes four Dirac fermions \cin{with two kinds of mass terms} at high-symmetry points,
$k^{*}_\alpha=(0,0), (0,\pi),(\pi,0),(\pi,\pi)$. 
In what follows, we set $\lambda_\mu=\lambda$ for simplicity. 
Around these points, the Hamiltonian is approximated by
\begin{alignat}1
{\cal H}_\alpha(k)/\lambda\equiv{\cal H}_\alpha(k)=\gamma^\mu_\alpha k_\mu+\gamma^{\mu+2}_\alpha m_{\mu,\alpha},
\label{DirBBH}
\end{alignat}
where $\mu=1,2$. \cin{This Hamiltonian has been referred to as the BBH Dirac insulator model.}
The subscript $\alpha$ of $\gamma$ matrices means that
not only the masses, but also $\gamma$-matrices depend on the symmetry points: 
$m_1=1\pm\frac{\gamma_1}{\lambda}$ and $\gamma^{1,3}=\pm\Gamma^{1,3}$ for $k^{*}_1=0,\pi$, and 
$m_2=1\pm\frac{\gamma_2}{\lambda}$ and $\gamma^{2,4}=\pm\Gamma^{2,4}$ for $k_2^{*}=0,\pi$.
As we can change the signs of any two of the $\Gamma$ matrices by unitary transformations, 
we can redefine each fermion ${\cal H}_\alpha$ with common $\gamma^\mu~(=\Gamma^\mu)$ matrices. 
The mass parameters are summarized in Table \ref{t:masses}.
We have to mention that the boundary matrices $S_j$ ($j=1,2$) introduced below are also independent of $k_\alpha^*$.
Thus, we will suppress $\alpha$, but we should keep it in mind that the mass terms are dependent on $k_\alpha^*$.

\begin{table}[h]
\begin{center}
\begin{tabular}{c|cccc}
\hline
$\alpha$ & $(0,0)$& $(\pi,0)$& $(0,\pi)$& $(\pi,\pi)$\\
\hline\hline
$m_1$& $1+\gamma_1$ & $1-\gamma_1$ & $1+\gamma_1$ & $1-\gamma_1$\\
$m_2$& $1+\gamma_2$ &$1+\gamma_2$ & $1-\gamma_2$& $1-\gamma_2$\\
\hline
\end{tabular}
\end{center}
\caption{
\cinb{
Mass parameters $m_1,\,m_2$ at four points $\alpha$. 
We have set $\lambda=1$.
}
}
\label{t:masses}
\end{table}


For the \cin{massive} Dirac fermion (\ref{DirBBH}), let us introduce a uniform magnetic field $B$ (around $\pi$ flux) in the $3$-direction 
(total magnetic flux per plaquette is $\pi+Ba^2$). Then, the Hamiltonian becomes 
\begin{alignat}1
{\cal H}&=-i\gamma^\mu D_\mu+\gamma^{\mu+2}m_\mu,
\label{BBHDirHam}
\end{alignat}
where $D_\mu=\partial_\mu-ieA_\mu$. 
\cin{This defines the BBH Dirac insulator model in a magnetic field.}
We expect it to be an effective model describing the properties \cin{as HOTI} of the lattice model with a magnetic flux around $\pi$ flux. 
In this paper, we choose the vector potentials in the Landau gauge such that
\begin{alignat}1
A_1=0,\quad A_2=Bx_1,
\label{LanGau}
\end{alignat}
to obtain explicit wave functions in the next section.

\subsubsection{Symmetries of the model}

\cin{Corresponding to Eq. (\ref{BBHMagSym}),} the \cin{BBH Dirac insulator} (\ref{BBHDirHam}) obeys the following transformation laws:
\begin{alignat}1
&M_1{\cal H}(x_1,x_2,B)M_1^{-1}={\cal H}(-x_1,x_2,-B),
\nonumber\\
&M_2{\cal H}(x_1,x_2,B)M_2^{-1}={\cal H}(x_1,-x_2,-B),
\nonumber\\
&T{\cal H}(x_1,x_2,B)T^{-1}={\cal H}(x_1,x_2,-B),
\end{alignat}
where $M_1$, $M_2$ are reflection matrices defined for the BBH model in Eq. (\ref{BBHSym}),
and $T=K$ denotes the time-reversal.
Define $\widetilde M_j=M_jT$ $(j=1,2)$. Then,  we have
\begin{alignat}1
&\widetilde M_1{\cal H}(x_1,x_2,B)\widetilde M_1^{-1}={\cal H}(-x_1,x_2,B),
\nonumber\\
&\widetilde M_2{\cal H}(x_1,x_2,B)\widetilde M_2^{-1}={\cal H}(x_1,-x_2,B),
\label{RefyMag}
\end{alignat}
The model in a magnetic field has also (antiunitary) reflection symmetries.

\cinb{
\subsubsection{Entanglement polarizations} \label{s:ep}
According to Ref. \cite{Fukui:2018aa}, we briefly discuss the topological invariants using the BBH Dirac fermion (\ref{BBHDirHam})
in the absence of a magnetic field.
Let $\psi(k)\equiv(\psi_1(k),\psi_2(k))$ be the ground state multiplet composed of two degenerate state,
and let $\rho(k)=\psi(k)\psi^\dagger(k)$ be the density matrix.
%
Introducing two kinds of partitions $(13)(24)$ and $(14)(23)$ within the unit cell in 
Fig. \ref{f:bbh},
we define the entanglement Hamiltonian ${\cal H}^{(13)}$ 
by tracing out $(24)$ degrees of freedom in the density matrix,
$\tr_{(24)}\rho(k)\propto e^{-{\cal H}^{(13)}}$, and likewise for
${\cal H}^{(14)}$.


Let us define two kinds of Berry connections $A^{(13)}(k)=\psi^{(13)\dagger}(k)\partial_{k_1}\psi^{(13)}(k)$
and $A^{(14)}(k)=\psi^{(14)\dagger}(k)\partial_{k_2}\psi^{(14)}(k)$, where $\psi^{(ab)}$ stands for the wave function
of the entanglement Hamiltonian ${\cal H}^{(ab)}$.
Integration of $A^{(13)}$ and $A^{(14)}$ over $k_1$ and $k_2$, respectively, yield entanglement polarizations 
$(p_1^{(13)},p_2^{(14)})$. For the continuum Hamiltonian (\ref{BBHDirHam}), 
they are given by 
$\left(\frac{1}{4}{\rm sgn~} m_1,\frac{1}{4}{\rm sgn~} m_2\right)$, which yields the invariant 
$q=2p_1^{(13)}p_2^{(14)}=\frac{1}{8}{\rm sgn~}m_1m_2$ for a single Dirac insulator.
Since the lattice model is composed of four fermions summarized in Table \ref{t:masses}, 
the topological invariant for the BBH model is 
\begin{alignat}1
q=\sum_\alpha q_\alpha.
\end{alignat}
This formula gives $q=\frac{1}{2}$ for $|\gamma_1|<1$ and $|\gamma_2|<1$, and $q=0$ for otherwise.
Thus, the Dirac fermion description of the BBH model reproduces the topological invariant for the lattice model,
taking doublers into account.


}

\subsubsection{Boundary conditions}\label{s:bbh_bou}

\cin{
In this subsection, we specify the boundary conditions of the BBH Dirac insulator when the model is defined on a half-plane.
Before discussing the boundary condition of the model, let us exemplify a boundary condition for 
a 1D tight-binding model with nearest neighbor hopping.
Let $H=\sum_{j=1}^\infty (c_{j+1}^\dagger t_j c_j+c_j^\dagger t_j^*c_{j+1}+c_j^\dagger v_j c_j)$ 
be a Hamiltonian defined on the semi-infinite
line $j\ge1$, and let $|\psi\rangle=\sum_{j=1}^\infty c_j^\dagger \psi_j|0\rangle$ be its eigenfunction.
The eigenvalue equation, $H|\psi\rangle=\varepsilon|\psi\rangle$,
is explicitly given by $t_{j-1}\psi_{j-1}+v_j\psi_j+t_j^*\psi_{j+1}=\varepsilon\psi_j$ $(j=1,2,\cdots)$.
When we consider the case $j=1$ of the above equation, 
it is natural to require $\psi_0=0$ as a boundary condition.
}
For the lattice BBH model in Fig. \ref{f:bbh}, 
we introduce a boundary between $j_1=0$ and $1$, and consider the system defined on the half-plane $j_1\geq1$.
Let $\psi_{jn}^\ell$ be the $n$th eigenstates of the Hamiltonian ${\cal H}^\ell_{ij}$ such that
${\cal H}^\ell_{ij} \psi_{jn}^\ell=\varepsilon_n\psi_{in}^\ell$.
Then, the boundary termination between $j_1=0,1$ is actualized by setting 
$\psi_{1,(j_1=0,j_2)n}^\ell=\psi_{4,(j_1=0,j_2)n}^\ell=0$.
Thus, the boundary condition of the lattice model is specified by
\begin{alignat}1
(S_1-1)\psi_{jn}^\ell\Big|_{j_1=0}=0, \quad S_1=-\tau^3\sigma^3=i\Gamma^1\Gamma^3.
\label{BBHLatBou}
\end{alignat}
Correspondingly, the same boundary condition should be imposed on the eigenstates of the continuum models such that
\begin{alignat}1
&(S_1-1)\psi_{n}(x)\Big|_{x_1=0}=0, \quad S_1=i\gamma^1\gamma^3.
\label{BBHDirBou}
\end{alignat}
Note here that $S_1$ does not depend on $k_\alpha^*$, as already mentioned.
Likewise, if one considers the system defined on $j_2\geq1$,
one can impose  the boundary condition on $j_2=0$, \cin{$\psi_{1,(j_1,j_2=0)n}^\ell=\psi_{3,(j_1,j_2=0)n}^\ell=0$.}
This is equivalent to impose the condition by using
$S_2=-\tau^0\sigma^3=i\Gamma^2\Gamma^4$ on the lattice wave function $\psi_{jn}^\ell$, 
and correspondingly, on the continuum wave function, 
\begin{alignat}1
(S_2-1)\psi_{n}(x)\Big|_{x_2=0}=0, \quad S_2=i\gamma^2\gamma^4.
\end{alignat}

\subsubsection{Symmetries of the boundary matrices}

So far, we have considered the system defined on $x_1\geq0$ imposing the boundary condition (\ref{BBHDirBou}).
If the system is defined on the opposite side $x_1\leq0$, the boundary condition is
\begin{alignat}1
&(S_1+1)\psi_{n}(x)\Big|_{x_1=0}=0.
\label{BBHDirBou2}
\end{alignat}
If the bulk system has reflection symmetry along the $x_1$ direction, two systems with a boundary at $x_1=0$,  
one defined on $x_1\ge0$ and the other defined on $x_1\le0$,  should be switched by reflection.
Here, note the following transformation laws of $S_1$:
\begin{alignat}1
M_1 S_1 M_1^{-1}=-S_1,\quad M_2 S_1M_{2}^{-1}=S_1.
\end{alignat}
The former ensures that the boundary conditions (\ref{BBHDirBou}) and (\ref{BBHDirBou2}) 
are indeed switched by reflection $M_1$. 
The latter relation means that the 
boundary condition in the $x_1$ direction is not affected by reflection $M_2$ in the $x_2$ direction.
Likewise, we have
\begin{alignat}1
M_1 S_2 M_1^{-1}=S_2,\quad M_2 S_2M_{2}^{-1}=-S_2,
\end{alignat} 
associated with reflection symmetry along the $x_2$ direction. 
Thus, the boundary conditions match the reflection symmetries.
Finally,
\begin{alignat}1
[S_1,S_2]=0,
\end{alignat}
implies that we can impose simultaneous boundaries both in the $x_1$ and $x_2$ directions.
This enables us to observe the corner states.

\subsubsection{Hermiticity of the Hamiltonian}

The BBH Dirac \cin{insulator} Hamiltonian Eq. (\ref{BBHDirHam}) should be Hermitian even with  a boundary 
\cite{1510.07698,Hashimoto:2016aa,Hashimoto:2017aa},
$\langle\phi|{\cal H}\psi\rangle=\langle{\cal H}\phi|\psi\rangle$.
Let us consider  the system  defined on the half-plane $x_1\geq0$. If we require 
\begin{alignat}1
(\widetilde S_1-1)\psi_{n}(x)\Big|_{x_1=0}=0, 
\end{alignat}
the Hamiltonian becomes Hermitian,
where $\widetilde S_1$ is any matrix satisfying $\{\widetilde S_1,\gamma_1\}=0$ and $\widetilde S_1^2=1$.
See discussions in Refs. \cite{1510.07698,Hashimoto:2016aa,Hashimoto:2017aa} and also in Sec. \ref{s:dir_nomag_edge}
in this paper.
Since $S_1$ defined in Eq. (\ref{BBHDirBou}) belongs to $\widetilde S_1$, 
the boundary condition (\ref{BBHDirBou}) due to the boundary termination of the
lattice model ensures the Hermiticity of the continuum BBH Dirac \cin{insulator} Hamiltonian. 
The Hermiticity in the $x_2$ direction is likewise.

\begin{figure*}[htb]
\begin{center}
\begin{tabular}{cccc}
\includegraphics[width=.23\linewidth]{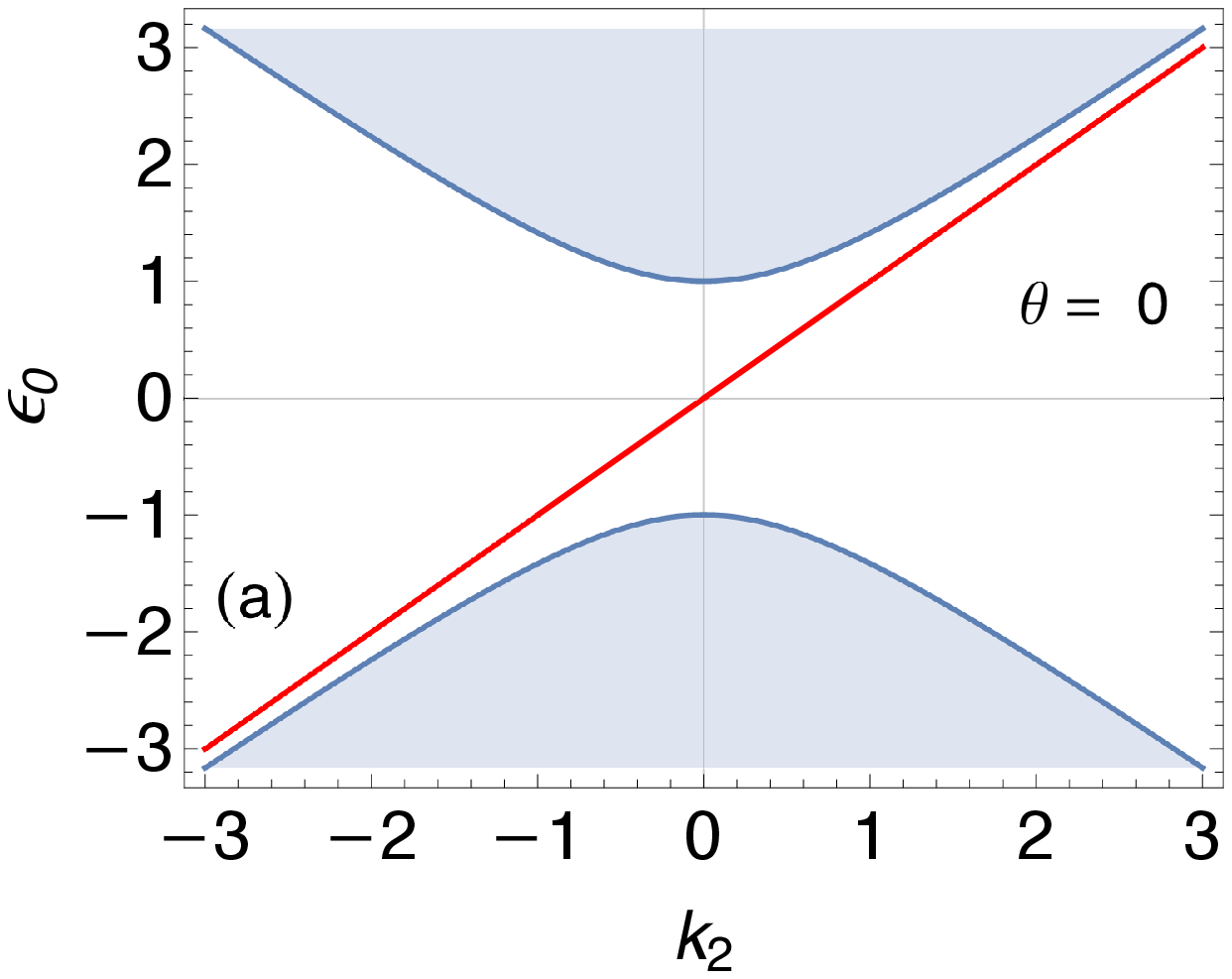}
&
\includegraphics[width=.23\linewidth]{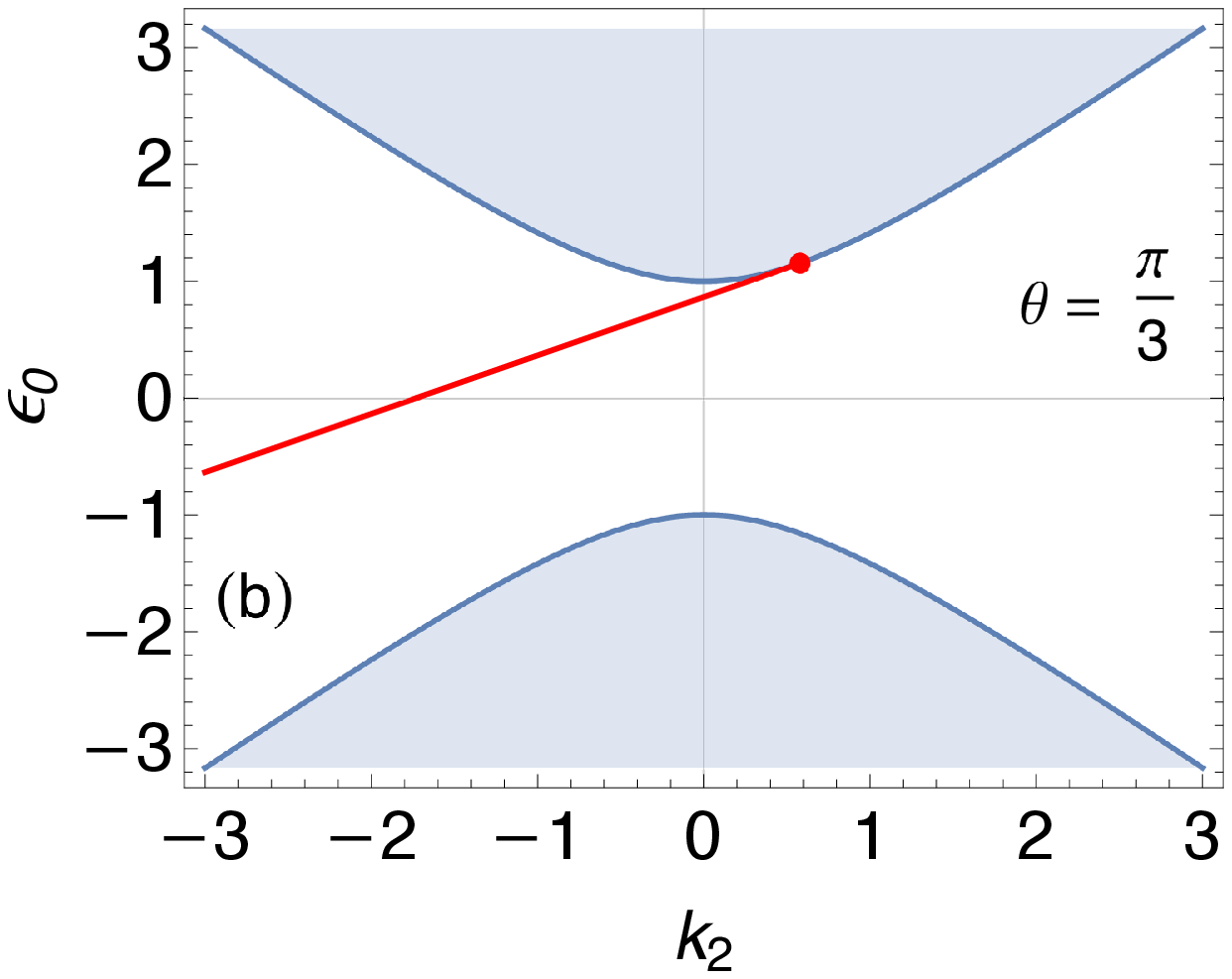}
&
\includegraphics[width=.23\linewidth]{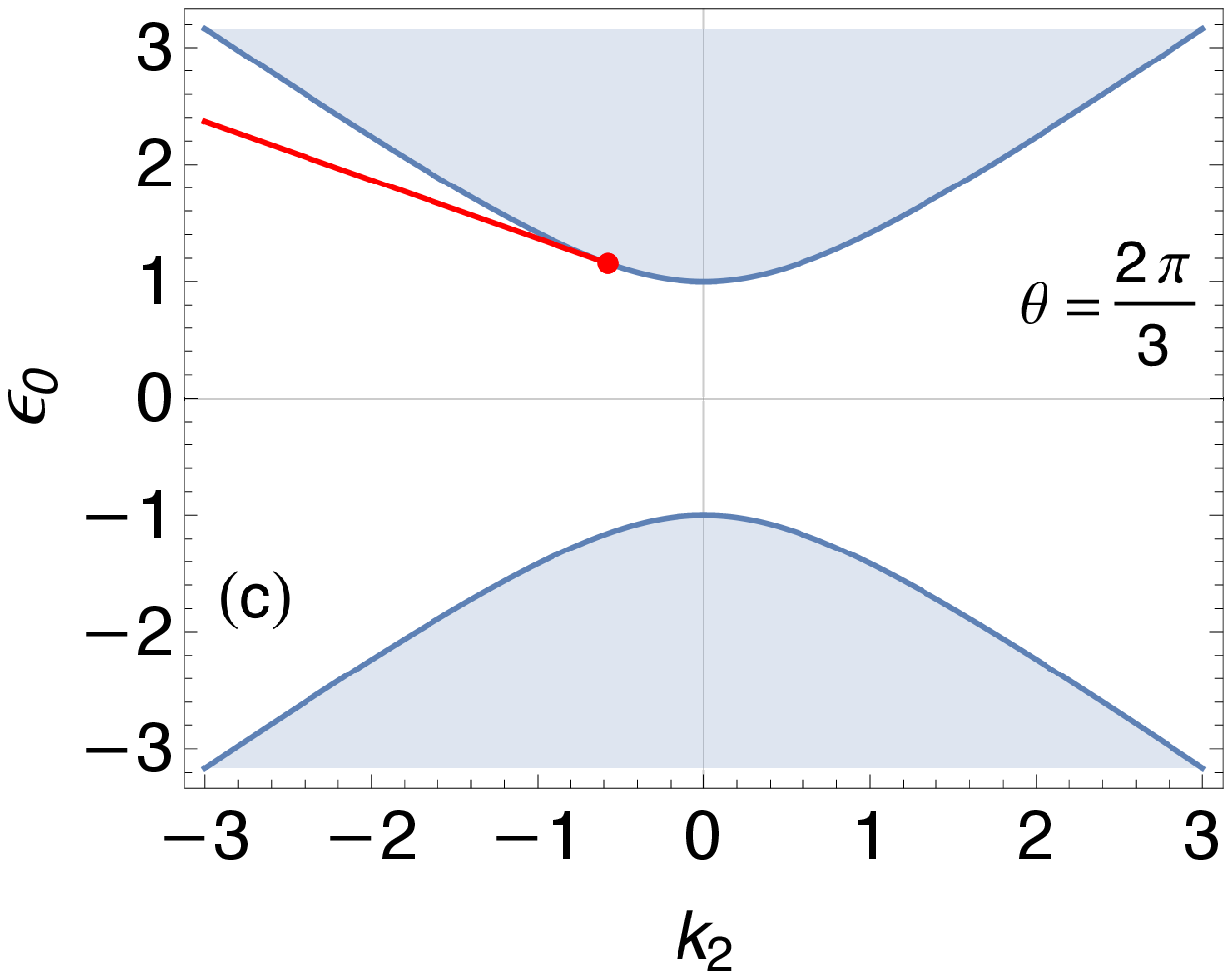}
&
\includegraphics[width=.23\linewidth]{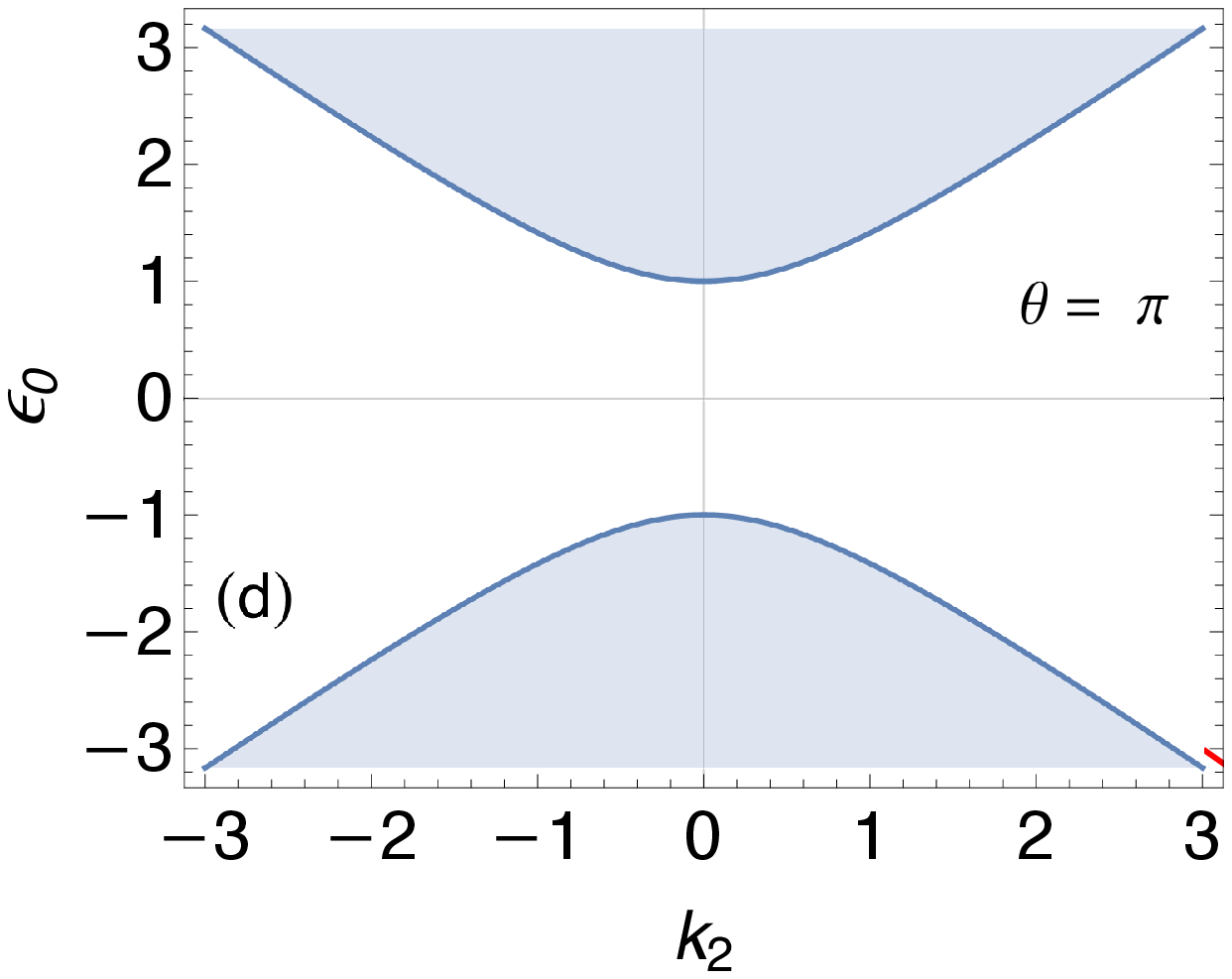}
\\
\includegraphics[width=.23\linewidth]{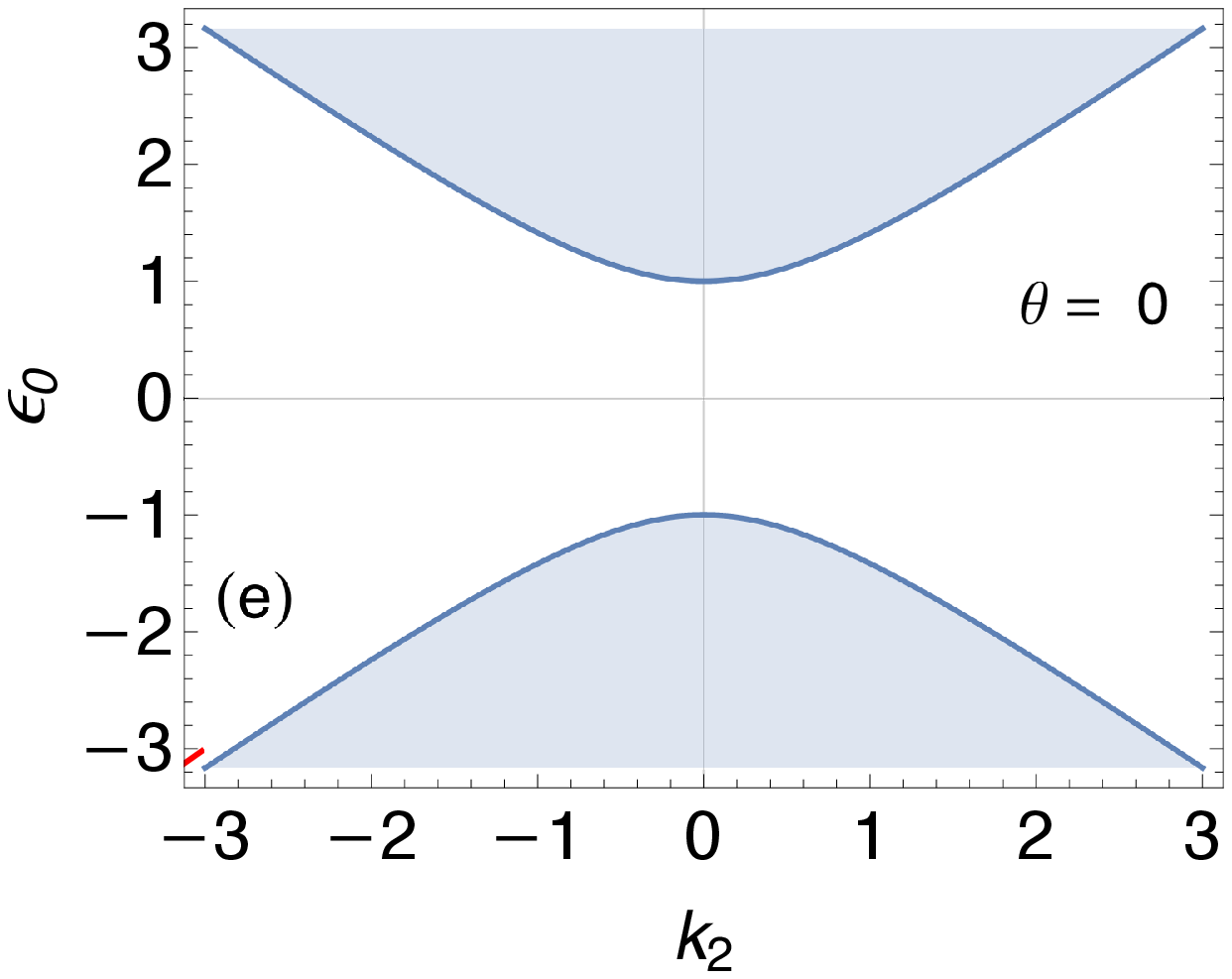}
&
\includegraphics[width=.23\linewidth]{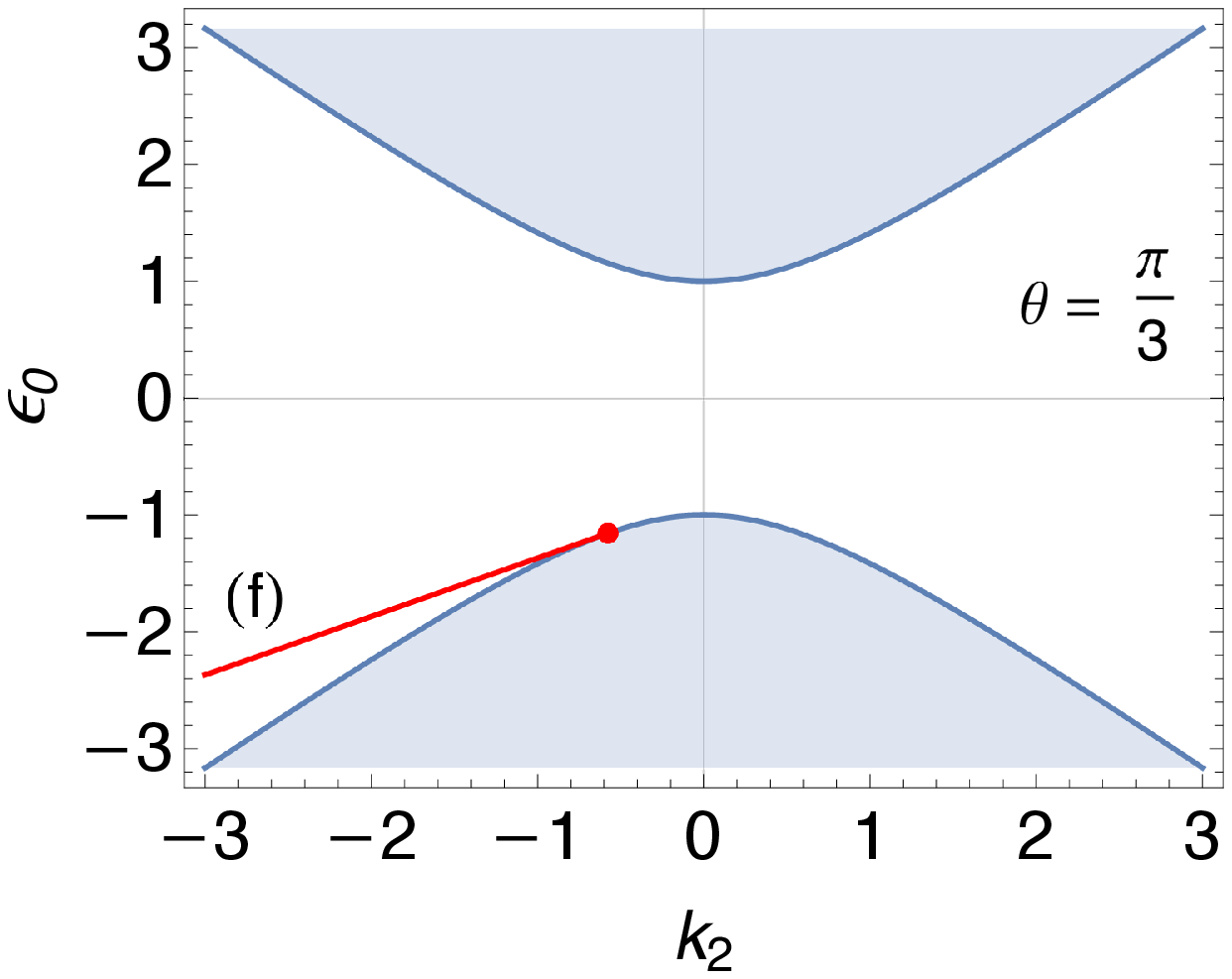}
&
\includegraphics[width=.23\linewidth]{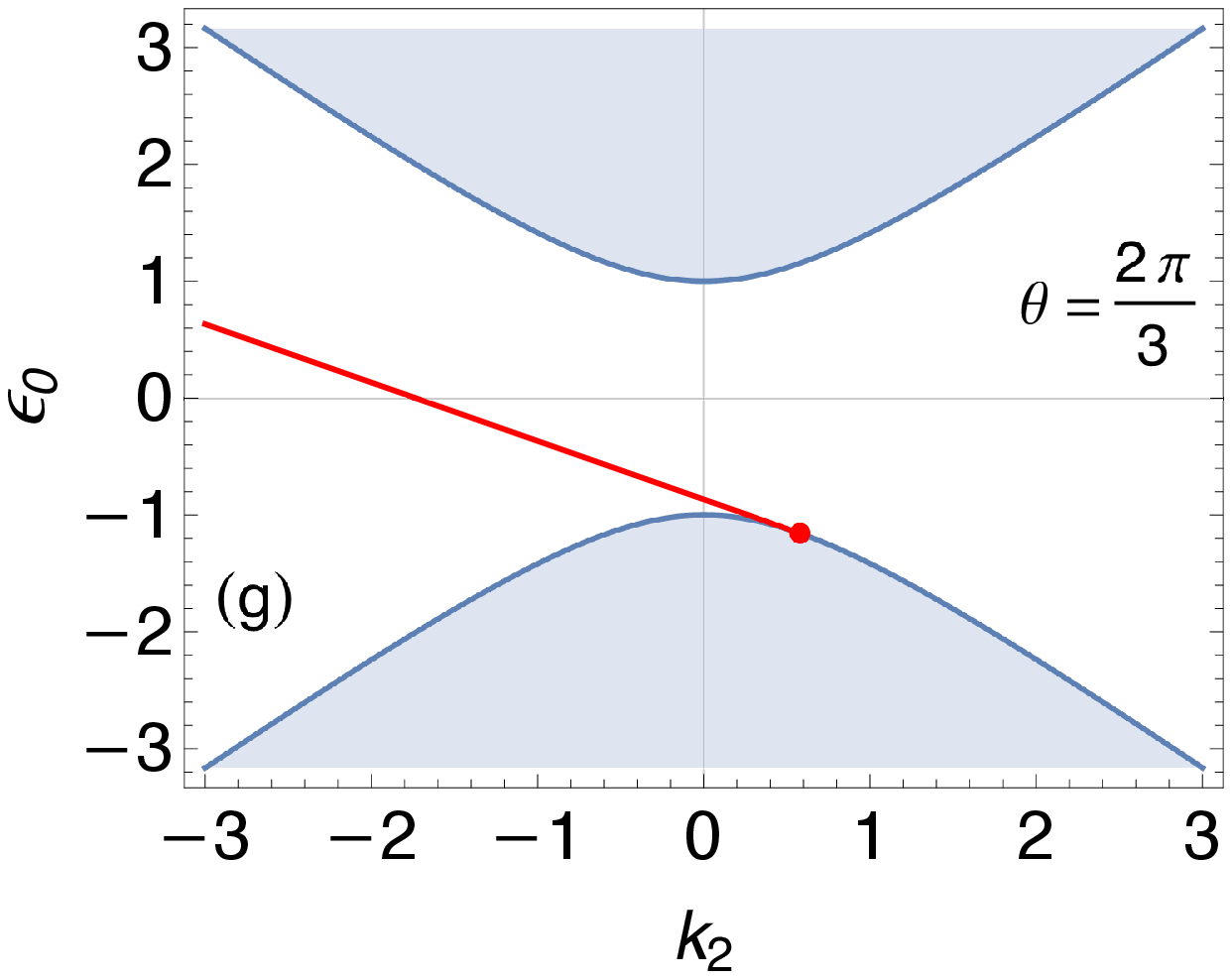}
&
\includegraphics[width=.23\linewidth]{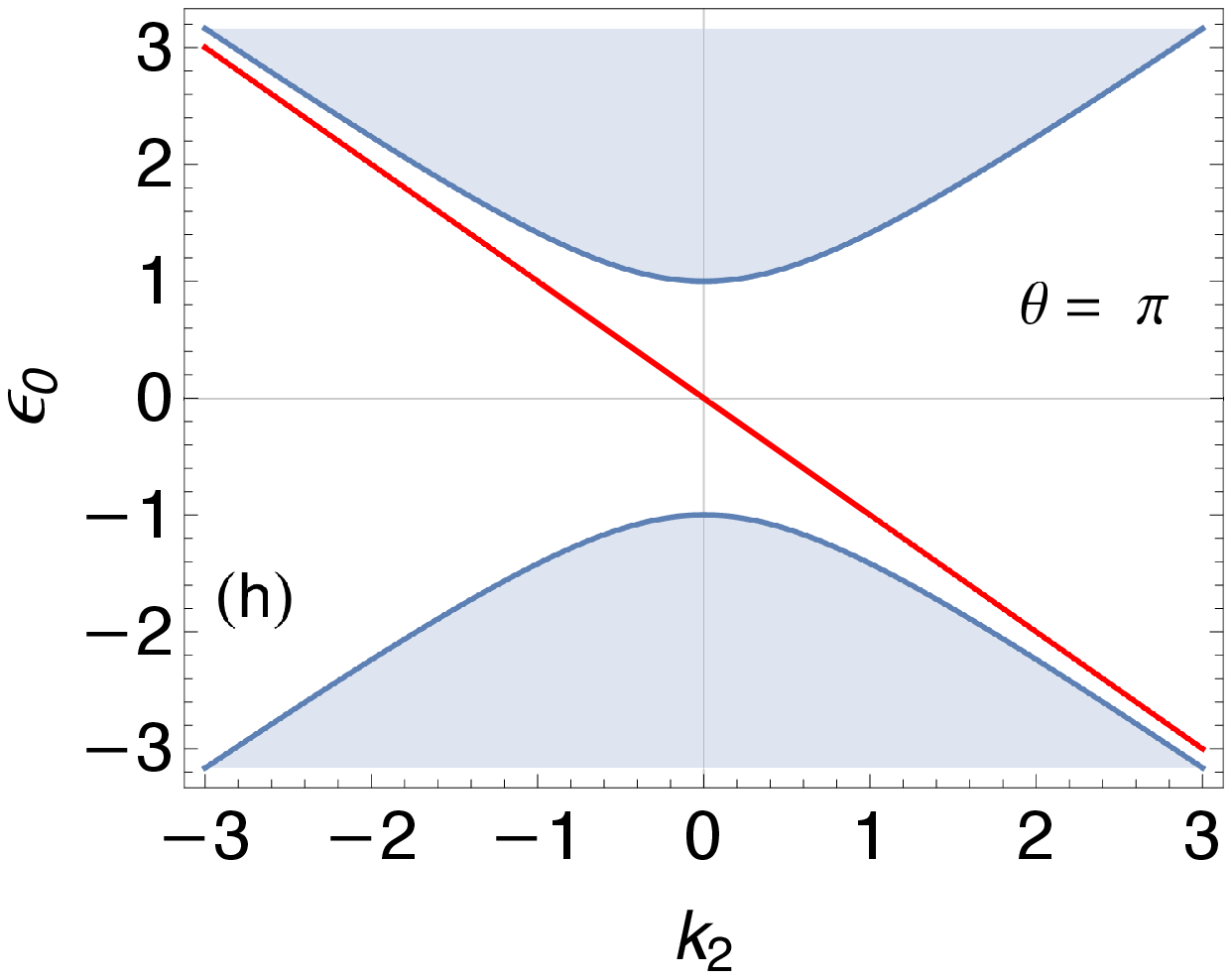}
\end{tabular}
\caption{
Edge state (red lines) obtained in Eqs. (\ref{WavFunPar}) and (\ref{RanK2})
with $m=1$ for upper four panels, whereas $m=-1$ for lower four panels.
Shaded regions denote the bulk spectra.
}
\label{f:edge}
\end{center}
\end{figure*}

\section{Conventional 2D Dirac insulator}\label{s:simple}

\cin{
This section is rather independent from other sections concerning the BOTP.
The motivation of this section is to derive edge states of the conventional massive Dirac fermion 
in a magnetic field. Here, by conventional massive Dirac fermion,
we mean a two-component fermion with a single mass term.
When we do not consider any corresponding lattice models, the guiding principle of the boundary conditions for 
the (massive) Dirac fermion may be the Hermiticity of the Hamiltonian, as studied in  
\cite{1510.07698,Hashimoto:2016aa,Hashimoto:2017aa} in the absence of a magnetic field.
In this section, we explore the theory of edge states for the massive Dirac fermion in the presence of
a magnetic field.

Basically, the Hermiticity conditions allow a parameter-dependence, so that we examine how 
edge states depend on such a parameter generically in this subsection, since it may clarify the 
relationship between edge states in the absence/presence of a magnetic field.
However, it should be stressed that tight-binding models on lattices choose an appropriate value of the parameter, 
as discussed in Sec. \ref{s:bbh_bou}.
Therefore, we will use only a specific boundary condition [$\theta=0$ in Eq. (\ref{BouCon})]
when we discuss the BOTP of the BBH Dirac insulator 
in Sec. \ref{s:bbh_dirac}.
}

Let us start with a conventional minimal 2D Dirac fermion with one mass term whose 
Hamiltonian is given by
\begin{alignat}1
{\cal H}_0=-i\sigma^\mu D_\mu+m\sigma^3  ,
\label{SimHam}
\end{alignat}
where $D_\mu=\partial_\mu-ieA_\mu$.
As shown in Sec. \ref{s:bbh_dirac}, edge states of the BBH Dirac \cin{insulator} can be derived by using those of Eq. (\ref{SimHam}).

\subsection{In the absence of a magnetic field}\label{s:dir_nomag}

\subsubsection{Bulk states}
The bulk Hamiltonian becomes
\begin{alignat}1
{\cal H}_0=\left(\begin{array}{cc} m&k_1-ik_2\\ k_1+ik_2&-m\end{array}\right) .
\end{alignat}
Therefore, the spectrum is given by
$\varepsilon_0(k)=\pm\sqrt{k^2+m^2}$ with $k^2=k_1^2+k_2^2$.

\subsubsection{Edge states}\label{s:dir_nomag_edge}

Assume that the system is defined on the half-plane $x_1\geq0$.
The Hamiltonian should be Hermitian, $\langle\phi|{\cal H}_0\psi\rangle=\langle{\cal H}_0\phi|\psi\rangle$
\cite{1510.07698,Hashimoto:2016aa,Hashimoto:2017aa}.
Form the integration by parts,
\begin{alignat}1
\int_0^\infty &dx_1\phi^\dagger(x)(-i\sigma^1\partial_1)\psi(x)=-i\phi^\dagger(x)\sigma^1\psi(x)\Big|_{x_1=0}
\nonumber\\
&+\int_0^\infty dx_1(-i\sigma^1\partial_1\phi)^\dagger(x)\psi(x),
\end{alignat}
we see that the Hermiticity of the Hamiltonian is ensured if the following condition is imposed:  
\begin{alignat}1
&(S_{1}^0(\theta)-1)\psi(x)\Big|_{x_1=0} =0,
\quad S_1^0(\theta)=\cos\theta \sigma^2+\sin\theta\sigma^3, 
\label{BouCon}
\end{alignat}
where $\theta$ is a fixed parameter, and $S_1^0(\theta)$ is a generic matrix that is anticommutative with $\sigma^1$
\cin{and satisfies $(S_1^0(\theta))^2=1$.}
Although $\theta$ is a free parameter for the continuum model, lattice models and their boundaries 
would choose a specific value of $\theta$,
as discussed in Sec. \ref{s:bbh_bou}.
In the absence of a magnetic field, the Hamiltonian Eq. (\ref{SimHam}) becomes
\begin{alignat}1
{\cal H}_0=
\left(\begin{array}{cc} m&-i\partial_1-ik_2\\ -i\partial_1+ik_2&-m\end{array}\right).
\label{FreDirEqu}
\end{alignat}
Let us solve ${\cal H}_0\psi_0(x_1,k_2)=\varepsilon_0\psi_0(x_1,k_2)$ for edge states.
Assume that 
\begin{alignat}1
\psi_0(x_1,k_2)=\frac{1}{\sqrt{\cal N}}\psi_0e^{iKx_1},\quad K=k_1+i\kappa,~(\kappa>0),
\label{DirWav}
\end{alignat}
where $\cal N$ is the normalization factor toward the $x_1$ direction. 
In what follows, such a normalization factor for wave functions will be suppressed, for simplicity.
Then, $\psi_0$ should be an eigenstate of $S_1^0(\theta)$: $S_1^0(\theta)\psi_0=\psi_0$, and hence,  
\begin{alignat}1
\psi_0=\left(\begin{array}{c} 1\\ \chi\end{array}\right),\quad \chi=\frac{\sin\theta-1}{i\cos\theta}.
\label{WavFunNoMag}
\end{alignat}
The eigenvalue equation becomes
\begin{alignat}1
\left(\begin{array}{cc} m&K-ik_2\\ K+ik_2&-m\end{array}\right)
\left(\begin{array}{c} 1\\ \chi\end{array}\right)=\varepsilon_0
\left(\begin{array}{c} 1\\ \chi\end{array}\right).
\end{alignat}
This equation leads to the following solutions for the edge state,
\begin{alignat}1
&\varepsilon_0=k_2 \cos\theta+m\sin\theta,
\nonumber\\
&k_1=0,\quad \kappa=-k_2\sin\theta+m\cos\theta(>0).
\label{WavFunPar}
\end{alignat}
The condition $\kappa>0~(e^{-\kappa}<1)$ restricts the range of $k_2$ such that
\begin{alignat}1
\begin{array}{cc}k_2<m\cot\theta \quad &(\sin\theta>0)\\k_2>m\cot\theta& (\sin\theta<0)\end{array}.
\label{RanK2}
\end{alignat}
In particular, when $\theta=0,\pi$, we have
\cin{
\begin{alignat}1
&\bullet\theta=0
\nonumber\\
&\quad\left\{\begin{array}{ll}
\varepsilon_0=k_2 , \psi_0(x_1,k_2)\propto e^{-mx_1}(1,i)^T&(m>0)\\
\mbox{no edge states}& (m<0)
\end{array}
\right. ,
\nonumber\\
&\bullet\theta=\pi,
\nonumber\\
&\quad\left\{\begin{array}{ll}
\mbox{no edge states}& (m>0)\\
\varepsilon_0=-k_2 , \psi_0(x_1,k_2)\propto e^{mx_1}(1,-i)^T&(m<0)
\end{array}
\right. .
\label{The0Edg}
\end{alignat}
}
Note that the edge states [Eq. (\ref{DirWav})] satisfy the boundary condition (\ref{BouCon}) not only 
at $x_1=0$ but also all along $x_1\geq0$.
On the other hand, as to the bulk states, not traveling waves $\psi_{0\pm} e^{\pm ik_1x}$ 
but their linear combination, i.e., the standing wave, can
satisfy the boundary condition (\ref{BouCon}) only at the boundary $x_1=0$. 
In Fig. \ref{f:edge}, we show some examples of the edge states obtained above.

\subsubsection{Effective Hamiltonian for  the edge state}\label{s:eff_edge_ham}

In the case with $\theta=0$, the effective Hamiltonian of the edge state becomes very simple.
The edge state obtained so far satisfies  Eq. (\ref{BouCon}) all along $x_1\geq0$.
Therefore, the edge state belongs to the subspace projected by $P=(1+S_x^0)/2=(1+\sigma^2)/2$.
Note that $P\sigma^1P=P\sigma^3P=0$. 
Thus, the effective Hamiltonian of the edge state toward the $x_2$ direction is given by
\begin{alignat}1
{\cal H}_{0,\rm 1edge}=P{\cal H}_0P=P(-i\partial_2)P.
\end{alignat}
Since in this subspace, $\sigma^2$ can be set $\sigma^2=1$, we obtain
\begin{alignat}1
{\cal H}_{0,\rm 1edge}=
\left\{
\begin{array}{cc}
-i\partial_2 \quad&  (m>0)\\
\mbox{no edge states}&(m<0)
\end{array}
\right. .
\label{EffEdgeHam}
\end{alignat}
This is of course consistent with the previous result in Eq. (\ref{The0Edg}).

\begin{figure*}[htb]
\begin{center}
\begin{tabular}{cccc}
\includegraphics[width=.23\linewidth]{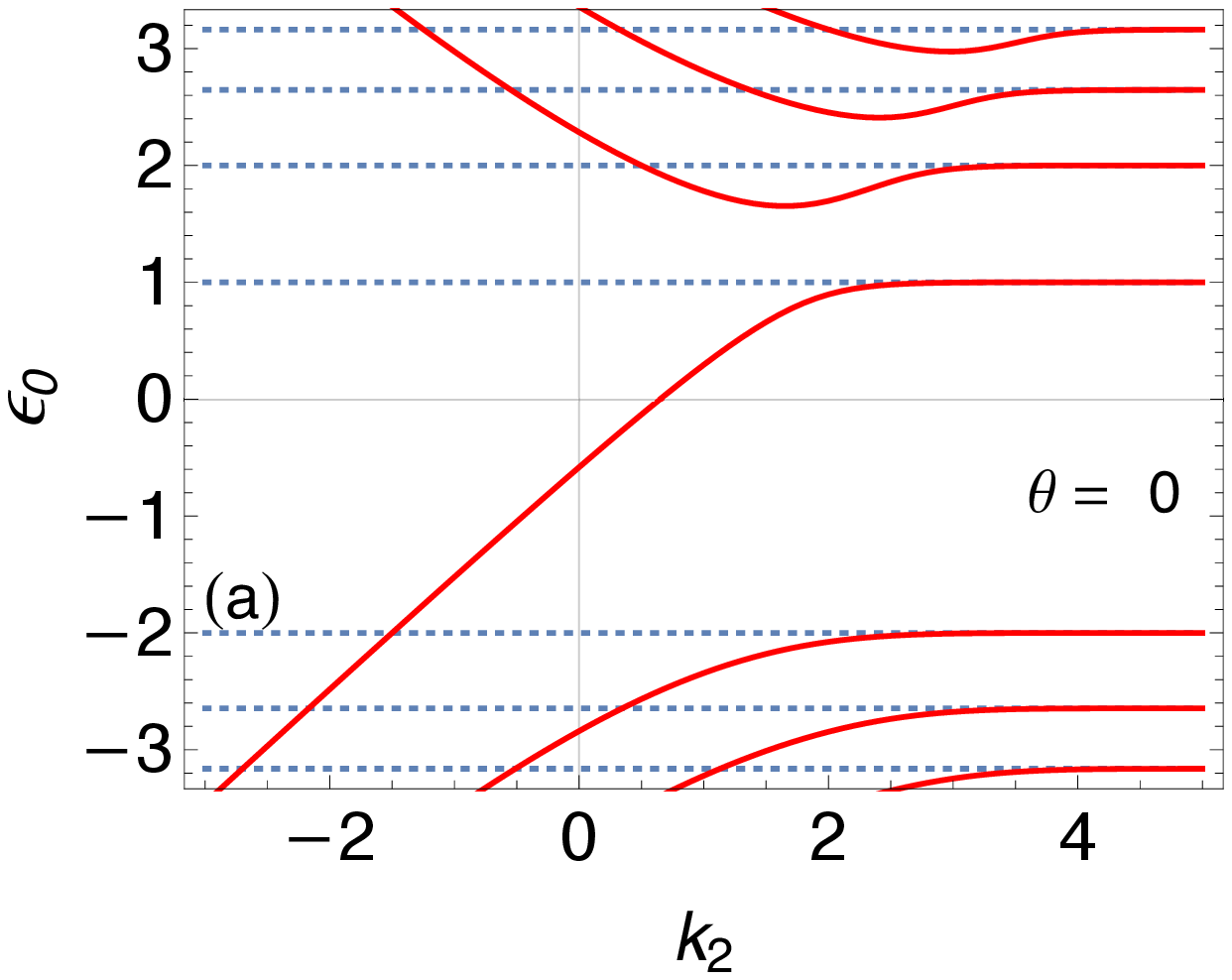}
&
\includegraphics[width=.23\linewidth]{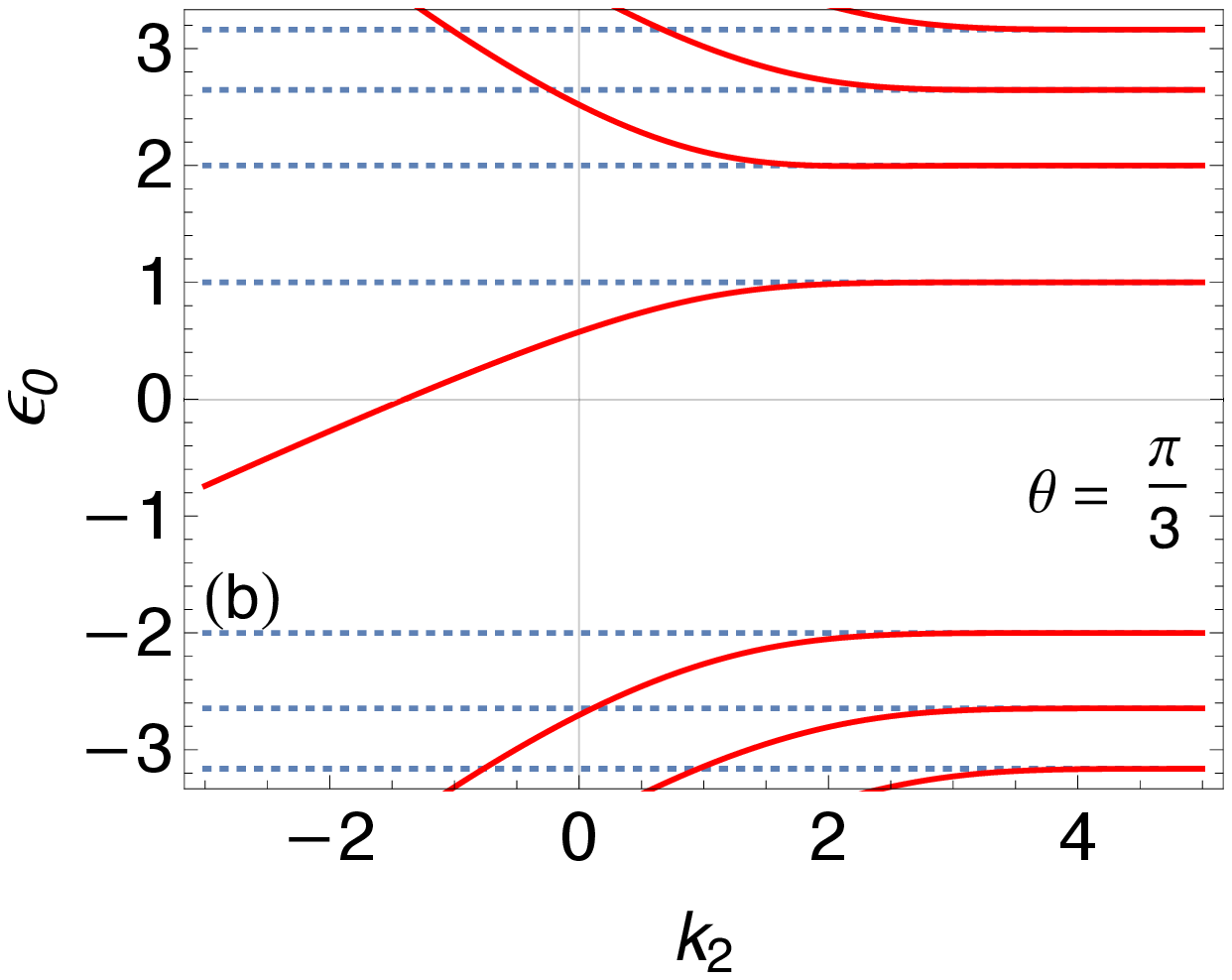}
&
\includegraphics[width=.23\linewidth]{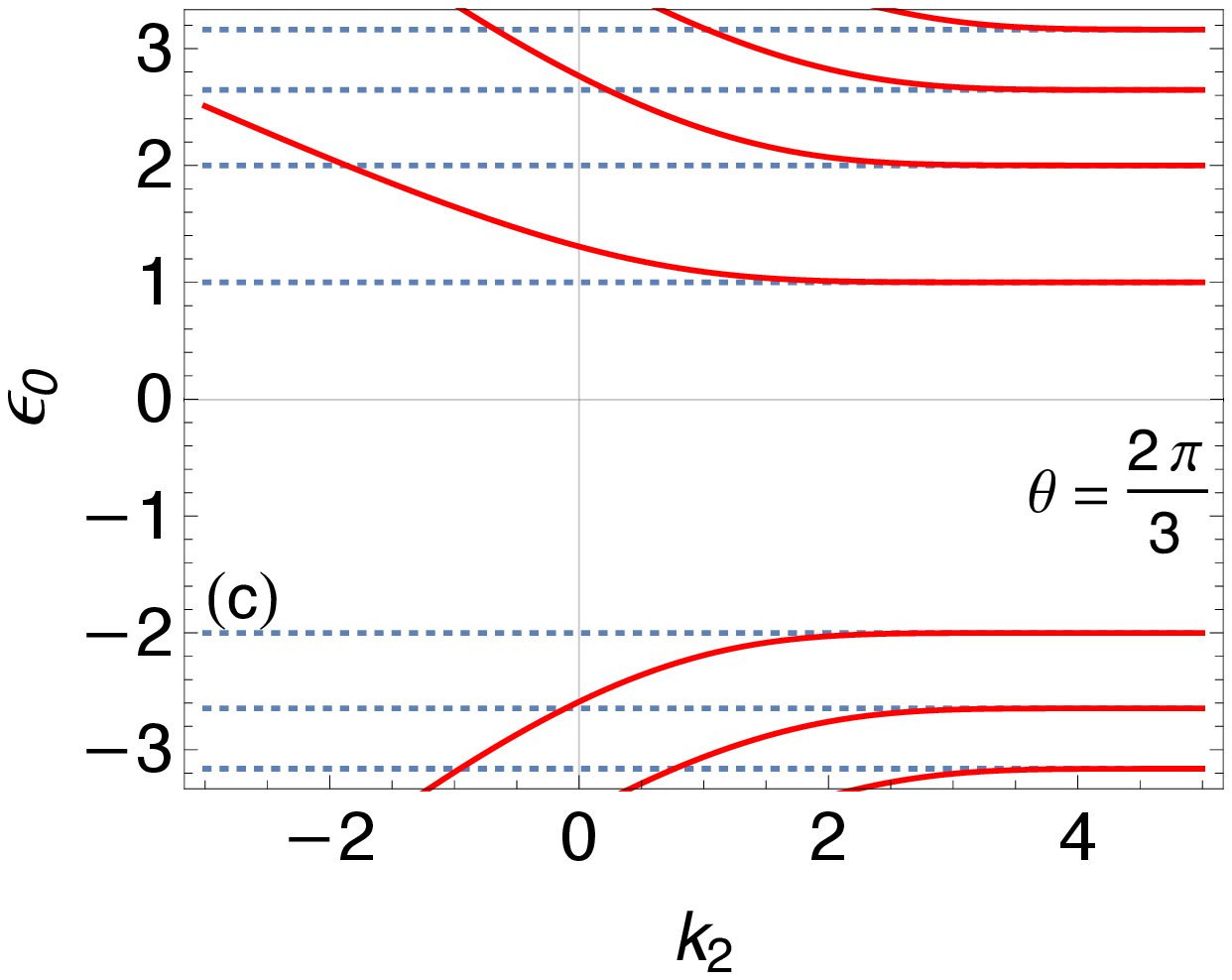}
&
\includegraphics[width=.23\linewidth]{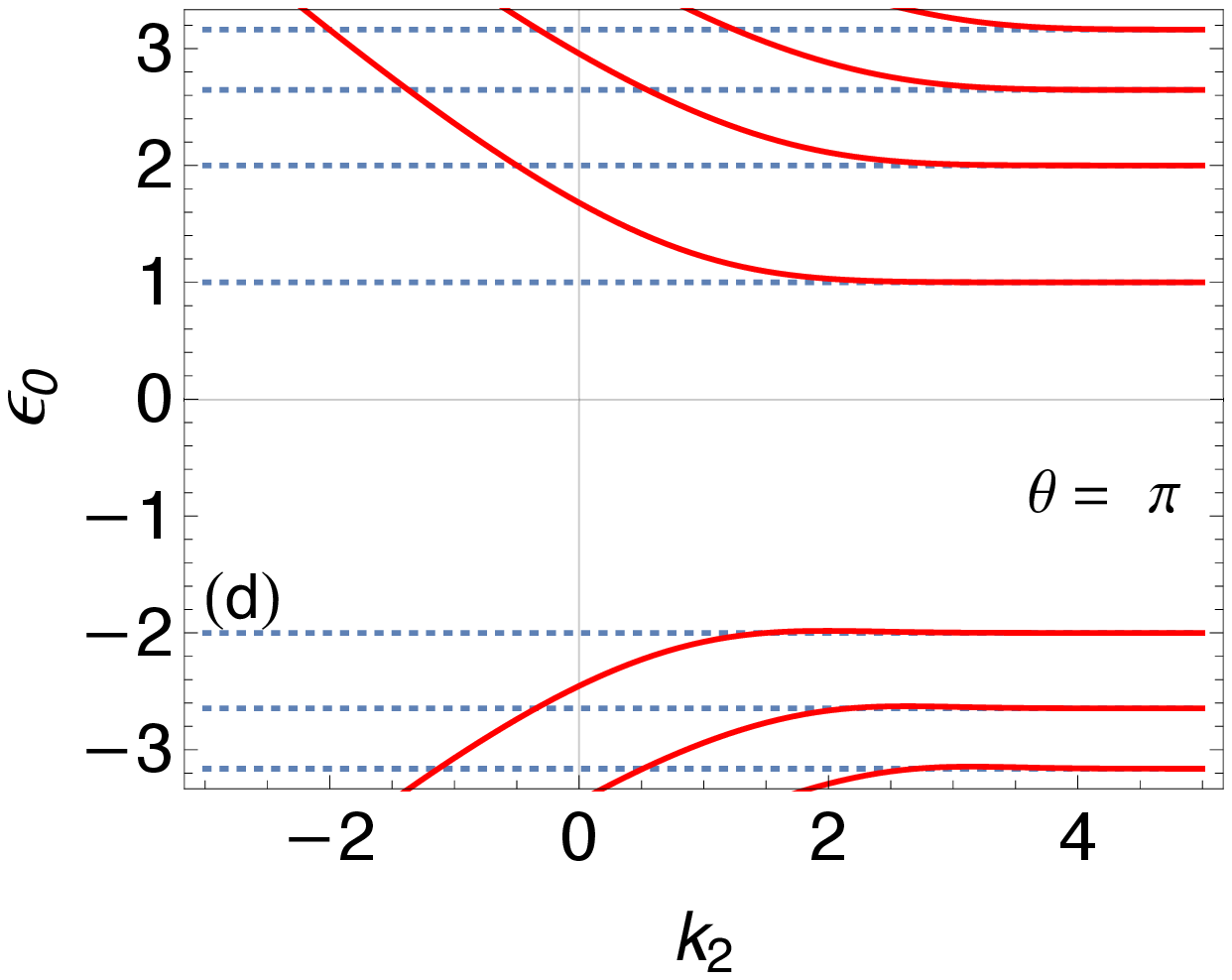}
\\
\includegraphics[width=.23\linewidth]{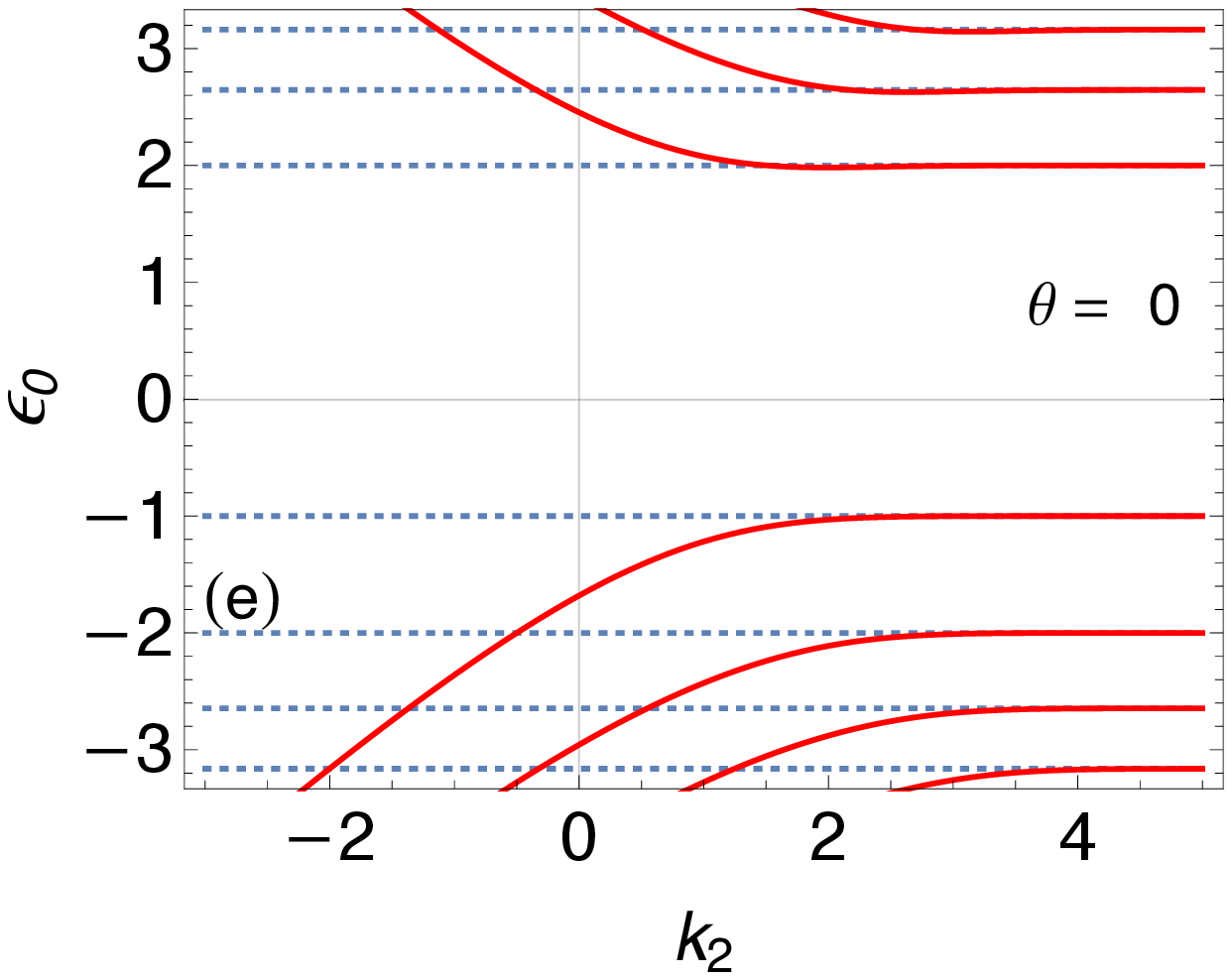}
&
\includegraphics[width=.23\linewidth]{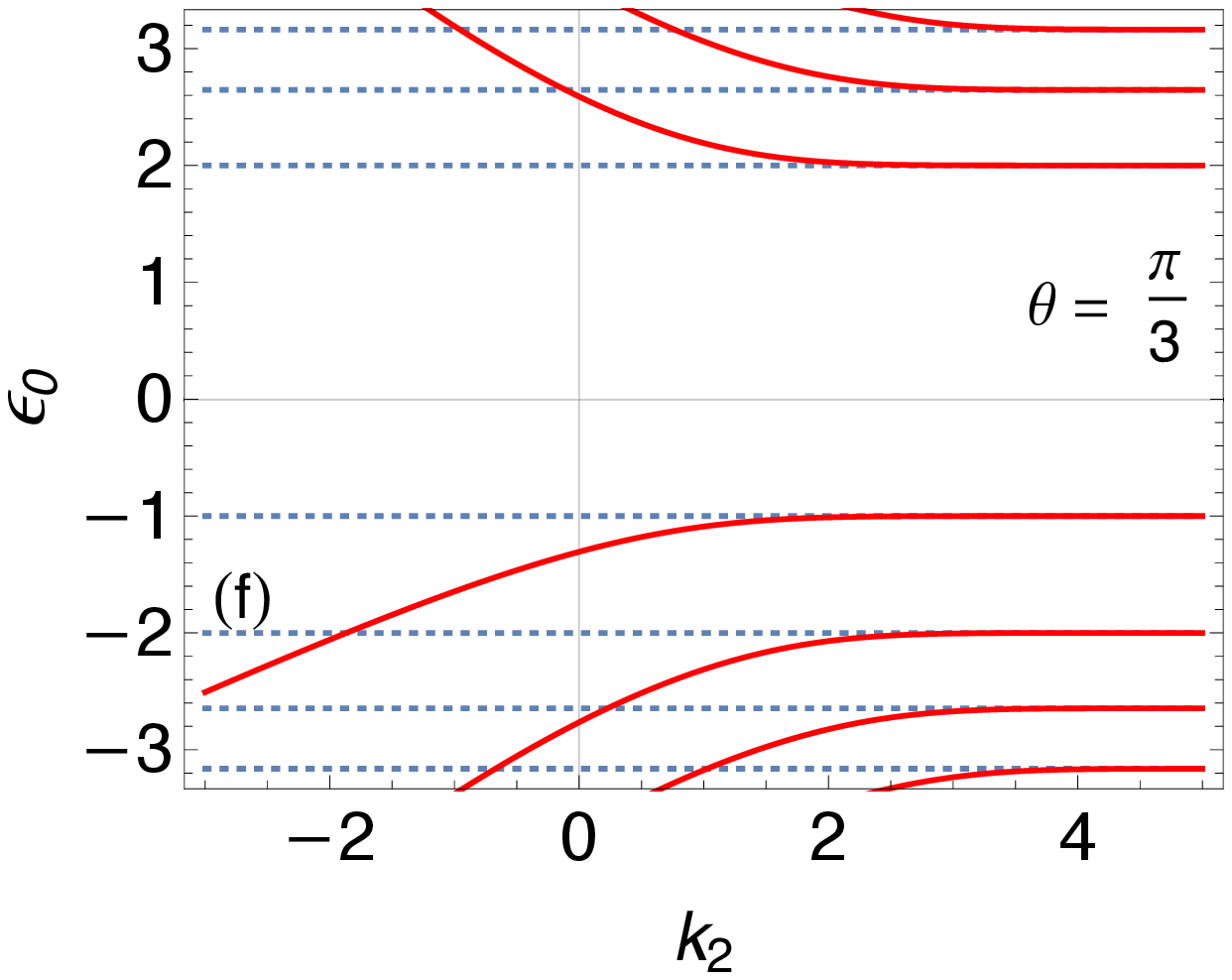}
&
\includegraphics[width=.23\linewidth]{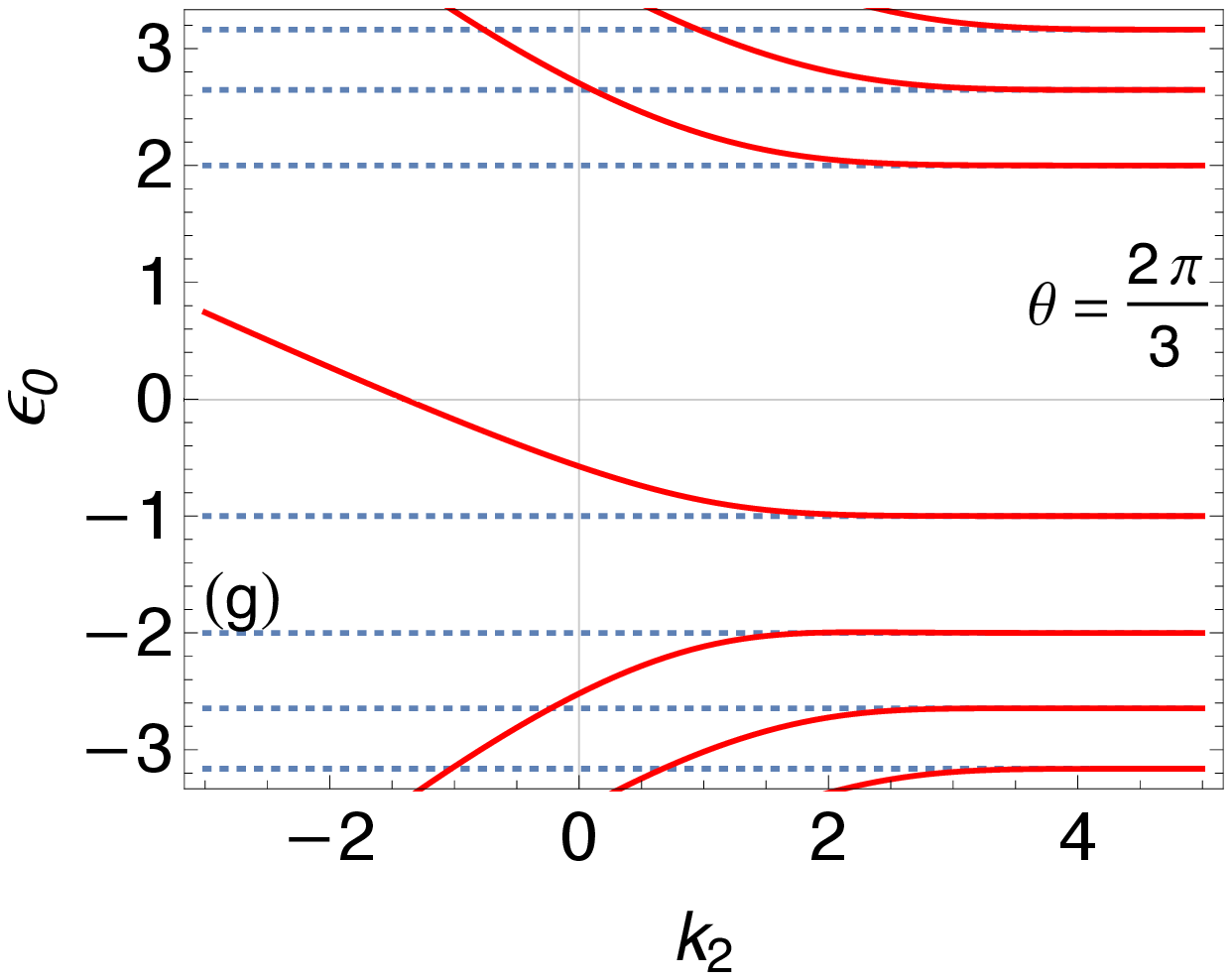}
&
\includegraphics[width=.23\linewidth]{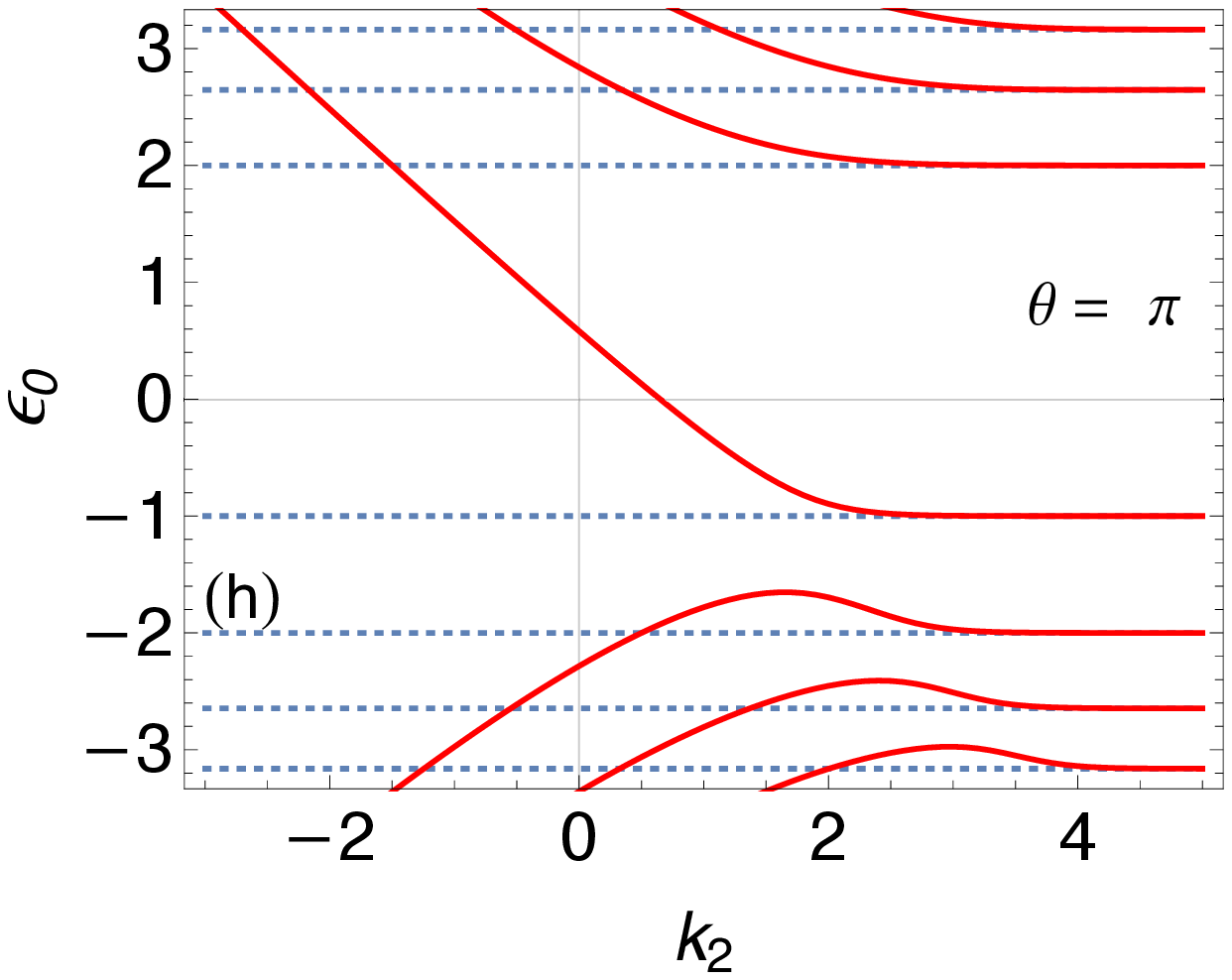}
\end{tabular}
\caption{
Edge (and bulk) states with $eB=1.5$ and $m=1$ for upper four panels for different $\theta$, 
whereas $m=-1$ for lower four panels.
The dashed lines indicate the Landau levels of the bulk states given by Eq. (\ref{MagP}).
}
\label{f:mag_edge}
\end{center}
\end{figure*}

\subsection{In the presence of a magnetic field}\label{s:simple_mag}

In this section, we derive edge states of the model (\ref{SimHam}) 
in the presence of  a uniform magnetic field.
We will show that the boundary condition (\ref{BouCon}) also plays a crucial role.

The Hamiltonian (\ref{SimHam}) becomes
\begin{alignat}1
{\cal H}_0=\left(\begin{array}{cc} m&-iD_1-D_2\\ -iD_1+D_2&-m\end{array}\right).
\label{DirHamMag}
\end{alignat}
It follows from 
$[D_1,D_2]
=-ieB$ 
that the following commutation relation holds,
\begin{alignat}1
[-iD_1+D_2,-iD_1-D_2]=2i[D_1,D_2]=2eB.
\end{alignat}
To obtain explicit wave functions, we choose the gauge potential given in Eq. (\ref{LanGau}).
Then, since the Hamiltonian does not depend on $x_2$, $-i\partial_2$ can be Fourier-transformed 
such that $-i\partial_2\rightarrow k_2$.
Therefore, we can define the creation and annihilation operators
\begin{alignat}1
& a=\frac{-iD_1+s D_2}{\sqrt{2|eB|}}=-i\left(\frac{d}{dz}+\frac{z-sz_0}{2}\right),  
\nonumber\\
&a^\dagger =\frac{-iD_1-s D_2}{\sqrt{2|eB|}}=-i\left(\frac{d}{dz}-\frac{z-sz_0}{2}\right),
\label{CreAni}
\end{alignat}
where $s=\mbox{sgn}\,eB$, $z=\sqrt{2|eB|}x_1$ and $z_0=\sqrt{\frac{2}{|eB|}}k_2$.
Using these operators, the Hamiltonian can be written as 
\begin{alignat}1
{\cal H}_0=
\left\{
\begin{array}{ll}
\left(\begin{array}{cc} m&\sqrt{2eB}a^\dagger\\ \sqrt{2eB}a&-m\end{array}\right)\quad&(s=1)\\
\left(\begin{array}{cc} m&\sqrt{2|eB|}a\\ \sqrt{2|eB|}a^\dagger&-m\end{array}\right)\quad&(s=-1)
\end{array}
\right.
\end{alignat}

Now we assume $eB>0$ ($s=1$), and solve the eigenvalue equation,
\begin{alignat}1
\left(\begin{array}{cc} m&\sqrt{2eB}a^\dagger\\ \sqrt{2eB}a&-m\end{array}\right)
\left(\begin{array}{c} \varphi\\\chi\end{array}\right)=
\varepsilon_0\left(\begin{array}{c} \varphi\\\chi\end{array}\right).
\label{DirEquMag}
\end{alignat}
The upper component $\varphi(z)$ obeys 
$(2eBa^\dagger a+m^2)\varphi(z)=\varepsilon_0^2\varphi(z)$, which can be written as, 
\begin{alignat}1
&\left(\frac{d^2}{dz^2}+\nu+\frac{1}{2}-\frac{1}{4}(z-z_0)^2\right)\varphi(z)=0,
\nonumber\\
&\nu=\frac{\varepsilon_0^2-m^2}{2eB},
\label{WeberEqu}
\end{alignat}
using Eq. (\ref{CreAni}).
It is known that the solution of Eq. (\ref{WeberEqu}) 
is given by $\varphi(z)=D_\nu(z-z_0)$, where $D_\nu(z)$ is 
the parabolic cylinder function \cite{high-tran-2}. 
Although 
this function is divergent at $z\rightarrow-\infty$ for generic $\nu$, as in Eq. (\ref{asy2}),
it is convergent at $z\rightarrow\infty$ and  normalizable on the semi-infinite line 
$0<z<\infty$, as in Eq. (\ref{asy1}). 
For several useful formulas of the parabolic cylinder functions, see Appendix \ref{s:app}.

Note that the lower component satisfies $\chi(z)=\sqrt{2eB}a\varphi(z)/(\varepsilon_0+m)$. Thus, 
the eigenfunction is given by
\begin{alignat}1
\psi_{0,\nu}(x_1,k_2)&\equiv
\left(\begin{array}{c} 1\\ \displaystyle\frac{\sqrt{2eB}a}{\varepsilon_0+m}\end{array}\right) D_\nu(z-z_0)
\nonumber\\
&=
\left(\begin{array}{c}  D_\nu(z-z_0)\\ -i\displaystyle\frac{\varepsilon_0-m}{\sqrt{2eB}}D_{\nu-1}(z-z_0)\end{array}\right),
\label{WavFunMag}
\end{alignat}
where we have used Eq. (\ref{CreAniWav}) and the normalization factor has been suppressed.
It follows from Eq. (\ref{WeberEqu}) that generically two paired eigenstates with opposite energies appear for a fixed $\nu$.
Likewise, in the case of $s=-1$,  we obtain
\begin{alignat}1
\psi_{0,\nu}(x_1,k_2)
&=
\left(\begin{array}{c} -i\displaystyle\frac{\varepsilon_0+m}{\sqrt{2|eB|}}D_{\nu-1}(z+z_0)\\D_\nu(z+z_0)\end{array}\right).
\label{WavFunMagNeg}
\end{alignat}
In what follows, we restrict our discussions to the case of $s=1$.

\subsubsection{Bulk states}

The bulk wave function should be normalized on the infinite line, $-\infty<z<\infty$. Therefore, $\nu$ is restricted to 
non-negative integers, $\nu=0,1,2,\cdots\equiv n$, and
eigenvalues and eigenfunctions are obtained, in the case of $s=1$, such that
\begin{alignat}1
\begin{array}{ll}
&\varepsilon_{0,0}=m,
\\
&\qquad\psi_{0,0}(x_1,k_2)=\left(\begin{array}{c} D_0(z-z_0)\\0\end{array}\right),
\\
&
\\
&\varepsilon^\pm_{0,n(>0)}=\pm\sqrt{2eBn+m^2},
\\
&\qquad \psi^\pm_{0,n}(x_1,k_2)=\left(\begin{array}{c}  D_n(z-z_0)\\    
-i\displaystyle\frac{\varepsilon^{\pm}_{0,n}-m}{\sqrt{2eB}}D_{n-1}(z-z_0) \end{array}\right).
\end{array}
\label{MagP}
\end{alignat}
These are famous Landau levels of a massive Dirac fermion \cite{Ishikawa:1985uq}.
When $m=0$, chiral symmetry ensures that the positive and negative levels are always paired 
except for zero energy. In the present model, there appears one zero-energy state.
When the mass becomes finite, the nonzero energy Landau levels are shifted in such a way that 
they are still paired in positive and negative energies. 
The zero-energy Landau level moves to energy $m$, 
and has no partner. This level causes the spectral asymmetry, which has intimate relationship
with the parity anomaly \cite{alvarez85,Ishikawa:1984aa,Semenoff:1984aa,Redlich:1984kx,Redlich:1984uq}
and is responsible for the bulk topological invariant \cite{Ishikawa:1985uq}.

\subsubsection{Edge states}\label{s:edge_and_bulk}

When the system is defined on $x_1\geq0$, 
the wave functions (\ref{WavFunMag}) or (\ref{WavFunMagNeg}) are always normalizable.
Instead of the nomalizability, the boundary condition (\ref{BouCon}) imposed on these wave functions
determines the eigenvalues and eigenstates.
To be concrete, the boundary condition on the wave function (\ref{WavFunMag}) is given by
\begin{alignat}1
&\sin\theta D_\nu(-z_0)-\cos\theta \frac{\varepsilon_0-m}{\sqrt{2eB}}D_{\nu-1}(-z_0)=D_{\nu}(-z_0),
\label{BouEqu}
\end{alignat}
where $\varepsilon_0$ and $z_0$ are defined, respectively, in Eq. (\ref{WeberEqu}) and below Eq. (\ref{CreAni}).
This is a nonlinear equation which determines $\varepsilon_0$ as a function of $k_2$.
It is not difficult to solve this equation using, e.g., {\it Mathematica} which includes parabolic cylinder functions 
as built-in functions.

We show in Fig. \ref{f:mag_edge}, numerical solutions of Eq. (\ref{BouEqu}) in the case of $eB>0$.
The eight panels in Fig. \ref{f:mag_edge} correspond to those in Fig. \ref{f:edge} in the absence of  a magnetic field. 
For large $k_2$, the spectra of Eq. (\ref{BouEqu}) converge to those of the bulk Landau levels (\ref{MagP}). 
This is natural, since the center of the harmonic potential in Eq. (\ref{WeberEqu}), $x_1=k_2/(eB)$,
is located far from the boundary at $x_1=0$ in the case of $s=1$. 
However, when $k_2$ becomes smaller, and at a certain value, 
$k_2\sim0$, 
boundary effects become larger and the states gradually change  their characters.
In this region, the spectra move away from those of the bulk Landau levels:
Basically, the positive and negative Landau levels go towards more positive and negative energies, respectively, 
when $k_2$ decreases from positive to negative values.
This is of course due to the boundary effects: Since Eq. (\ref{WeberEqu}) is the Schr\"odinger equation for the 
1D harmonic oscillator, one can expect that the boundary effects make $\varepsilon_0^2$ larger if $\varepsilon^2_0>m^2$.
However, the exception is the unpaired Landau level with energy $m$. 
As can be seen from the leftmost upper panel in Fig. \ref{f:mag_edge}(a) ($m>0$ and $\theta=0$), the $\varepsilon_0=m>0$
Landau level causes the spectral flow across zero energy. This level passes through the energies prohibited for the bulk system.
This is the edge state corresponding to the case in Sec. \ref{s:dir_nomag_edge} in the absence of a magnetic field. 
Indeed, the behavior of this edge state depends on $\theta$, and it resembles that in Fig. \ref{f:edge} as a function of $\theta$.
In particular,  $k_2\ll0$, they asymptotically become the same linear dispersions.

\begin{figure}[htb]
\begin{center}
\begin{tabular}{cc}
\includegraphics[width=.49\linewidth]{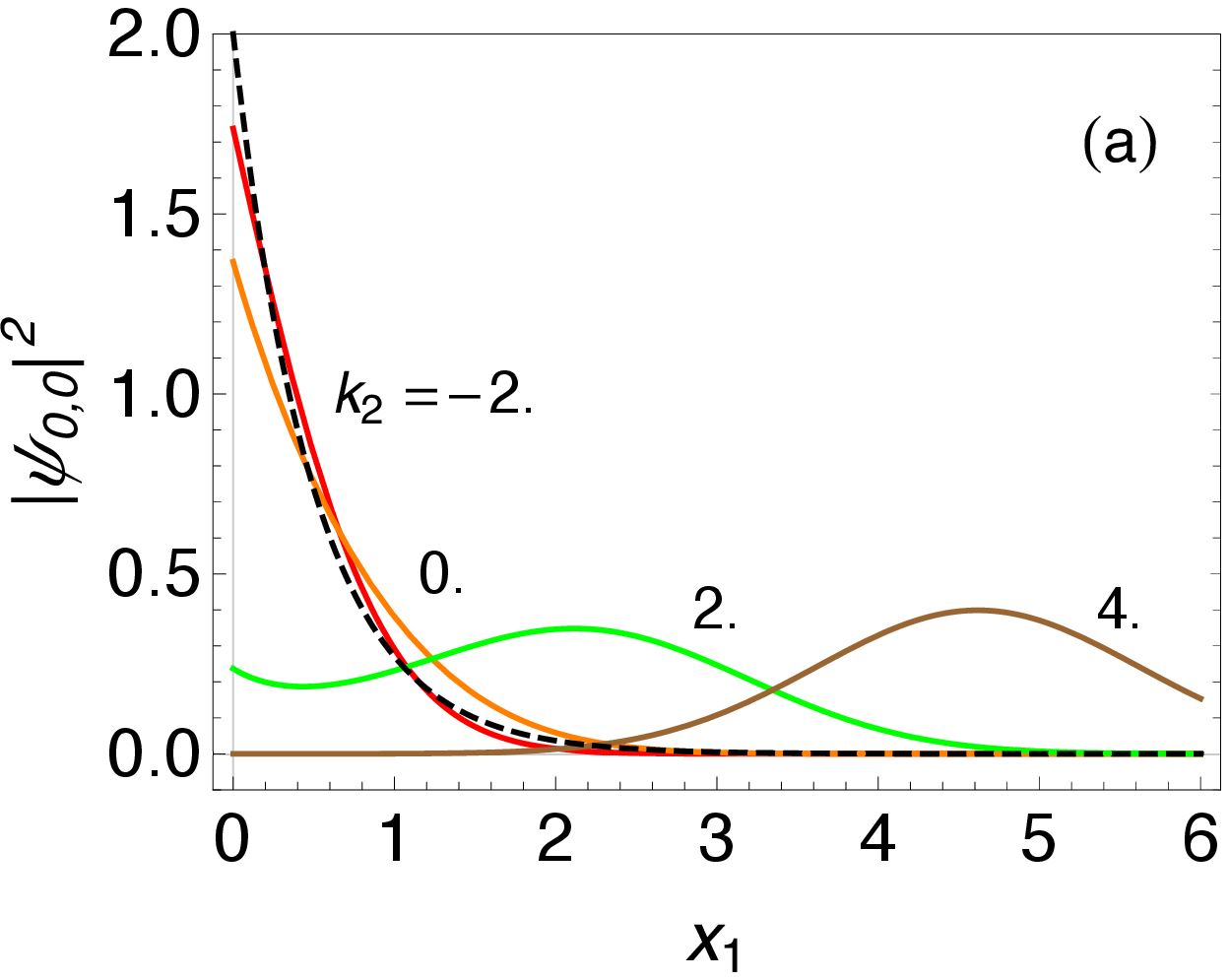}
&
\includegraphics[width=.49\linewidth]{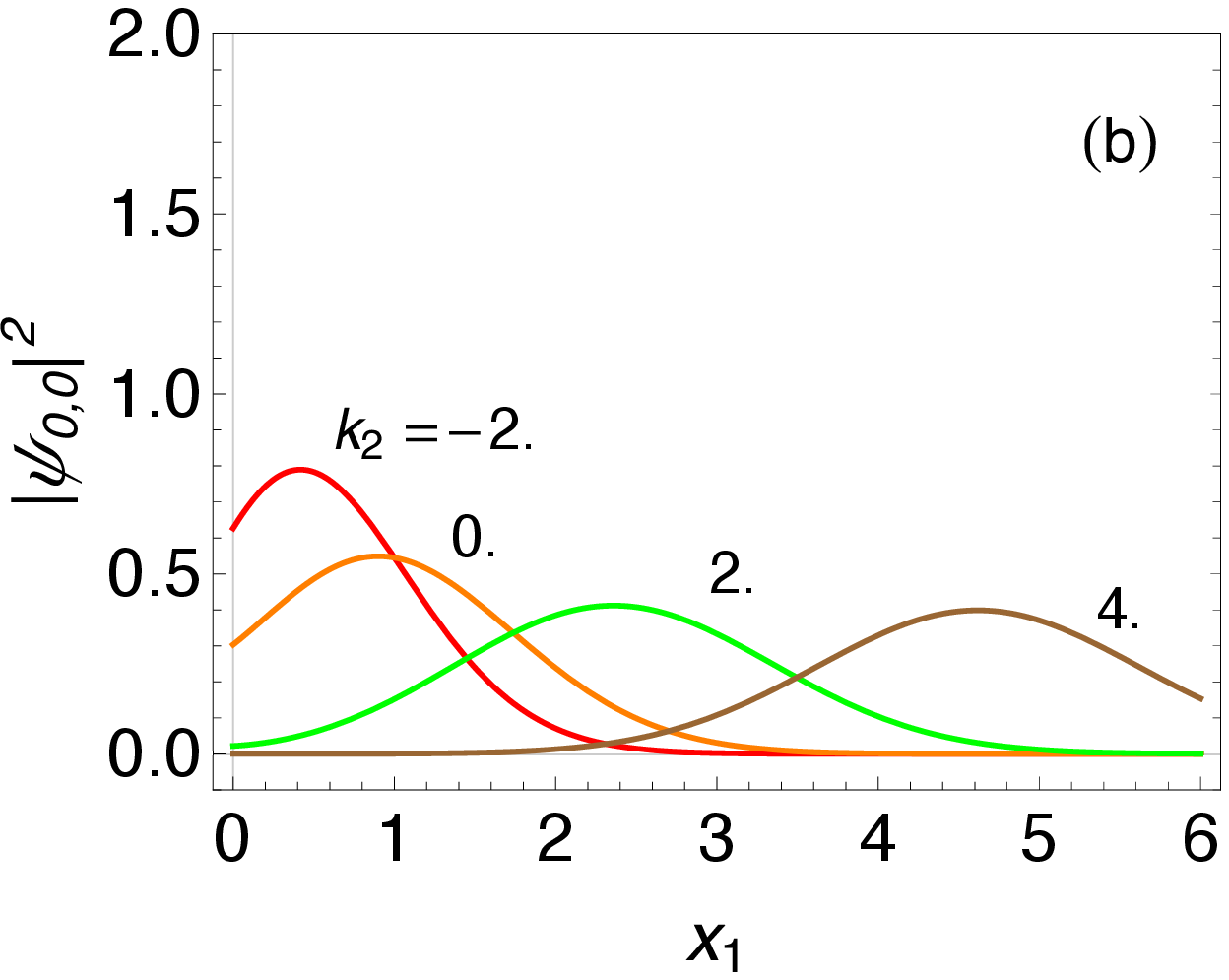}
\end{tabular}
\caption{
Local density profile of the states of the unpaired Landau level, $|\psi_{0,n=0}(x_1,k_2)|^2$, for the
$\theta=0$ and $B=1.5$ system. (a) For $m=1$ [Fig. \ref{f:mag_edge}(a)] and (b) for $m=-1$ [Fig. \ref{f:mag_edge}(e)].
Solid colored curves denote the density profiles at different $k_2$ values, whereas the dashed curve in (a) denotes
the density profile of edge state in the absence of magnetic fields $|\psi_0(x_1)|^2$ given in Eq. (\ref{DirWav})
which is independent of $k_2$ when $\theta=0$.
}
\label{f:profile}
\end{center}
\end{figure}

As shown in Sec. \ref{s:dir_nomag_edge},
the wave function $\psi_0(x_1,k_2)$ in Eq. (\ref{DirWav})  in the absence of a magnetic field
satisfies the boundary condition (\ref{BouCon}) not only at  $x_1=0$, but everywhere on $x_1\geq0$.
This enables us to obtain the effective Hamiltonian for the edge state in Sec. 
\ref{s:eff_edge_ham}.
Unfortunately, 
the wave functions in Eqs. (\ref{WavFunMag}) or (\ref{WavFunMagNeg}) satisfy the boundary condition 
only at the boundary, $x_1=0$, in the presence of a magnetic field. 

Therefore, to check the properties of the edge state in the presence of a magnetic field,
we compare the wave function of the unpaired Landau level $\psi_{0,n=0}(x_1,k_2)$ with
the exact wave function of the edge state $\psi_0(x_1,k_2)$ given by Eq. (\ref{The0Edg})
in the absence of a magnetic field.
In Fig. \ref{f:profile}, we show the local density profile 
for several $k_2$.
Figure \ref{f:profile} (a) is the case of  
the unpaired Landau level in Fig. \ref{f:mag_edge} (a). 
In this panel, we find that
when $k_2$ varies from positive to negative values, the wave function $\psi_{0,0}(x_1,k_2)$ changes its character from 
the bulk state in Eq. (\ref{MagP}) to the edge state in Eq. (\ref{DirWav}).

On the other hand, in the case of \cin{$m=-1$} in Fig. \ref{f:profile}(b), the local density has different profile from 
the edge state in Fig. \ref{f:profile}(a) even for negative $k_2$. The density profile is for bulk states rather than edge states.
Indeed, in this case, there are no edge states in the absence of magnetic field, as shown in Eq. (\ref{The0Edg}) as well as
in Fig. \ref{f:edge} (e).
Therefore, 
we conclude that {\it \cin{in the case of $\theta=0$}, the system with $m>0$ shows the edge state, as in Fig. \ref{f:mag_edge} (a), 
which can be approximated, for $k_2\lesssim 0$, by the  
edge state in the absence of a magnetic field in Fig. \ref{f:edge} (a), whereas the system with $m<0$ shows no edge state,
as in Fig. \ref{f:mag_edge} (e), which corresponds to Fig. \ref{f:edge} (e). }
Therefore, the effective Hamiltonian of the edge state in $k_2\lesssim0$ in the presence of a magnetic field is basically
given by Eq. (\ref{EffEdgeHam}) in the absence of a magnetic field.

\section{BBH Dirac insulator model}\label{s:bbh_dirac}

Based on the edge states derived so far, we discuss those of the BBH Dirac \cin{insulator} in Eq. (\ref{BBHDirHam}).
To this end, the following $\gamma$ matrices are convenient: 
\begin{alignat}1
&\gamma^j=\tau^1\sigma^j, ~(j=1,2,3),~
\gamma^4=\tau^2\sigma^0,~\gamma_5=\tau^3\sigma^0.
\label{NewGam}
\end{alignat}
Then, the Hamiltonian becomes
\begin{alignat}1
{\cal H}=
\left(
\begin{array}{cc}
&{\cal H}_0-im_2\\{\cal H}_0+im_2
\end{array}
\right),
\label{BBHHam}
\end{alignat}
where ${\cal H}_0$ is given by Eq. (\ref{SimHam}) with $m=m_1$.
\cin{This is the merit of using the $\gamma$ matrices in Eq. (\ref{NewGam}): The Hamiltonian of the BBH Dirac insulator 
is simply expressed by the  Hamiltonian studied in Sec. \ref{s:simple}, and, hence, the edge states derived there are 
directly used in the following discussions. In particular, in the discussion of the edge states in Sec. \ref{s:simple_mag}
in the presence of a magnetic field,
the Hamiltonian in the form of Eq. (\ref{DirHamMag}) is very convenient to rewrite the Hamiltonian with respect to 
the creation and annihilation operators.}
\cin{On the other hand,  }the boundary matrix $S_1$ in Eq. (\ref{BBHDirBou})  is $S_1=i\gamma^1\gamma^3=\tau^0\sigma^2$ in the 
present basis, \cin{although it is diagonal in the basis in Sec. \ref{s:bbh_bou}.
Thus, the wave functions of the edge states become a bit more complicated.}
In the present new basis Eq. (\ref{NewGam}), 
$S_1$ can be written as
\begin{alignat}1
S_1=\left(
\begin{array}{cc}S_1^0(0)&\\ &S_1^0(0)\end{array}
\right), 
\end{alignat}
where $S_1^0(0)$ is defined in Eq. (\ref{BouCon}) with $\theta=0$.
In what follows, we solve edge states for a half-plane $x_1\geq0$ satisfying
\begin{alignat}1
{\cal H}\psi(x)=\varepsilon\psi(x), \quad (S_1-1)\psi(x)\Big|_{x_1=0}=0.
\label{BBHDirEqu}
\end{alignat} 

\subsection{In the absence of a magnetic field}

\subsubsection{Bulk states}\label{s:BBHnomag}

The Hamiltonian in the momentum representation is
\begin{alignat}1
{\cal H}(k)=\gamma^\mu k_\mu+\gamma^{\mu+2}m_\mu.
\end{alignat}
Therefore, the bulk spectrum is 
\begin{alignat}1
\varepsilon(k)=\pm\sqrt{k^2+m^2}, 
\end{alignat}
where $k^2=k_1^2+k_2^2$ and $m^2=m_1^2+m_2^2$. 
Each state above is doubly degenerate. The bulk gap closing occurs at $m_1=m_2=0$ only.

\subsubsection{Edge states}\label{s:BBHnomagedge}

Let us consider the system defined on $x_1\geq0$.
Let $\psi_0(x_1,k_2)$ be the edge state wave function (\ref{DirWav}) of ${\cal H}_0$ in Eq. (\ref{FreDirEqu})
, i.e., ${\cal H}_0\psi_0(x_1,k_2)=\varepsilon_0\psi_0(x_1,k_2)$ 
satisfying the boundary condition (\ref{BouCon}) with $\theta=0$.
Then, for $m_1>0$
\begin{alignat}1
\psi(x_1,k_2)=\left(\begin{array}{l}\varepsilon\psi_0(x_1,k_2)\\(\varepsilon_0+im_2)\psi_0(x_1,k_2)\end{array}\right),
\label{BBHWav}
\end{alignat}
is the wave function satisfying Eq. (\ref{BBHDirEqu}).
Here, \cin{$\psi_0(x_1,k_2)=e^{-m_1x_1}(1,i)^T$ and $\varepsilon_0=k_2$ are the wave function and 
dispersion given in Eq. (\ref{The0Edg}),} and 
$\varepsilon=\pm\sqrt{\varepsilon_0^2+m_2^2}=\pm\sqrt{k_2^2+m_2^2}$ is the dispersion 
of the edge states \cin{of ${\cal H}$ in Eq. (\ref{BBHHam}).}
The gap closing of these edge states occurs at $m_2=0$ regardless of $m_1(>0)$.
On the other hand, when $m_1<0$, there are no edge states.  \cin{See Eq. (\ref{The0Edg}).}
Taking account of the discussions in Sec. \ref{s:eff_edge_ham},  an effective Hamiltonian of the 1D edge state 
localized along $x_1\sim0$ is 
given by
\begin{alignat}1
& m_1>0,
\quad{\cal H}_{\rm 1edge}=
\left(\begin{array}{cc}&-i\partial_2-im_2\\-i\partial_2+im_2\end{array}\right),
\nonumber\\ 
& m_1<0,
\quad\mbox{no edge states} .
\label{BBHEdgEff}
\end{alignat}
The above Hamiltonian ${\cal H}_{1\rm edge}$ is nothing but the 1D massive Dirac fermion.
\cinb{ 
This fermion yields the polarization $\frac{1}{4}{\rm sgn \,}m_2$ toward the $x_2$ direction. 
Therefore, at $m_2=0$, i.e., at $\gamma_2=\pm1$, topological 
changes occur.
} 
It should be noted that $\sigma^2$ of $S_1=\tau^0\sigma^2$ acts as $1$ in this subspace.

In addition to the boundary along $x_1=0$, 
let us introduce another boundary along $x_2=0$ and consider the above edge state (\ref{BBHEdgEff})
in the region $x_2\geq0$. 
As discussed in Sec. \ref{s:BBH}, the boundary condition toward $x_2$ for the BBH Dirac \cin{insulator} is given by 
$S_2=i\gamma^2\gamma^4=-\tau^3\sigma^2$ in the present basis (\ref{NewGam}).
In the subspace of Eq. (\ref{BBHEdgEff}), we can set $\sigma^2=1$, so that $S_2$ acts as
$S_2=-\tau^3$ in the space of Eq. (\ref{BBHEdgEff}).  
Thus,  the zero-dimensional edge state of ${\cal H}_{1\rm edge}$, 
\begin{alignat}1
&{\cal H}_{1\rm edge}\psi'(x_2)=\varepsilon'\psi'(x_2),
\nonumber\\
&(S_2-1)\psi'(x_2)\Big|_{x_2=0}=0,\quad S_2=-\tau^3,
\end{alignat}
is obtained as follows:
\begin{alignat}1
&m_2>0, \quad
\varepsilon'=0,\,\psi'(x_2)=\sqrt{2m_2}\left(\begin{array}{c}0\\e^{-m_2 x_2}\end{array}\right),
\nonumber\\
&m_2<0,\quad\mbox{no edge states}.
\label{BBHCor}
\end{alignat}
The above transition at $m_2=0$ with keeping $m_1>0$ is the boundary obstruction of the edge states
mentioned below Eq. (\ref{BBHWav}).
\cin{Thus, we have shown that the corner state exists in the case of $m_1,m_2>0$, whose wave function is given by
$\psi(x_1,x_2)\propto e^{-(m_1x_1+m_2x_2)}(0,0,1,i)^T$.}


\subsection{In the presence of a magnetic field}\label{s:bbh_mag}

Finally, we consider the BBH Dirac \cin{insulator} in a magnetic field in Eq. (\ref{BBHHam}).
In this section, we restrict our discussions to the case of $eB>0$.
Even in the presence of a magnetic field,  
completely the same discussions as Sec. \ref{s:BBHnomagedge} are applied to this case.

\begin{figure}[htb]
\begin{center}
\begin{tabular}{cc}
\includegraphics[width=.49\linewidth]{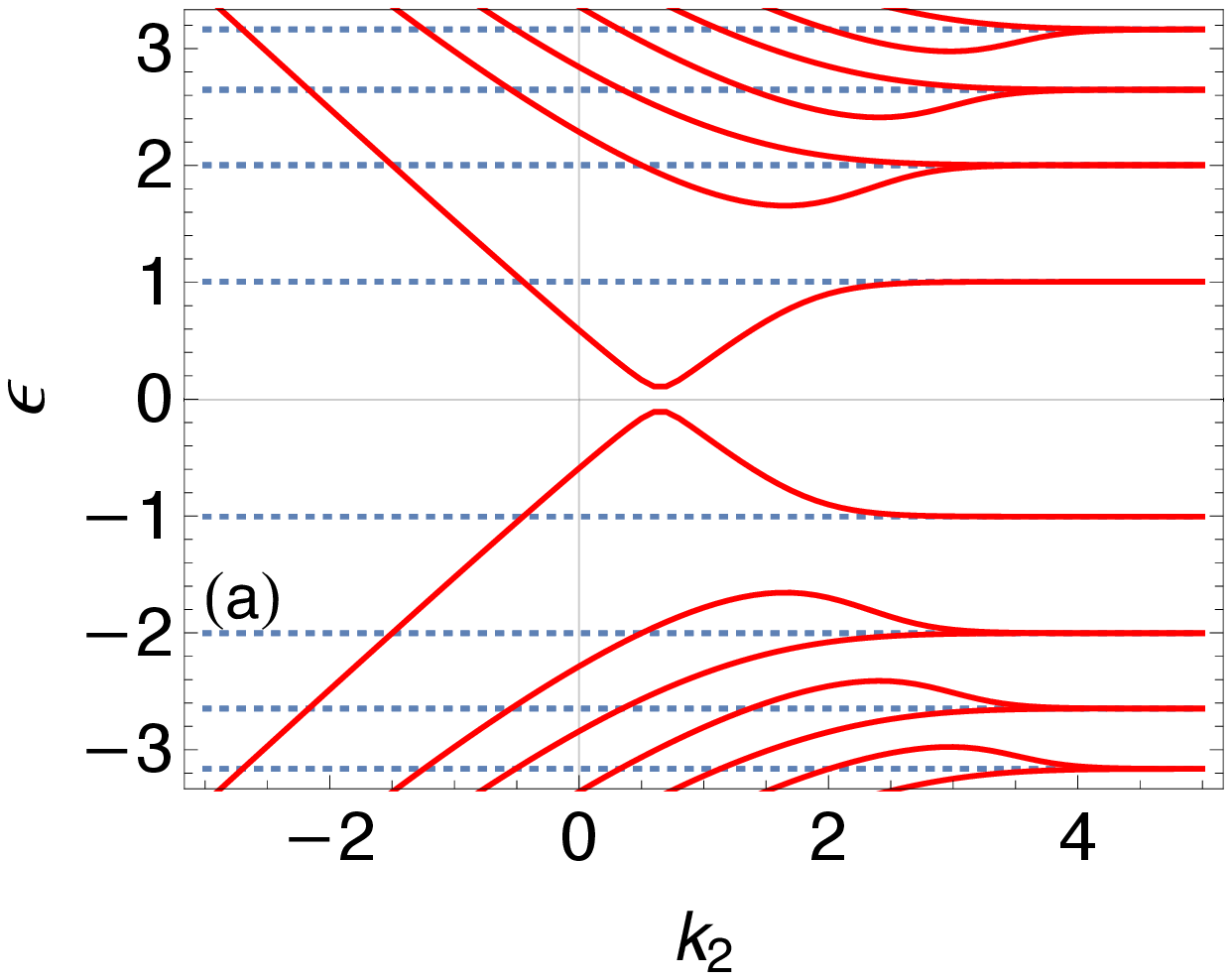}
&
\includegraphics[width=.49\linewidth]{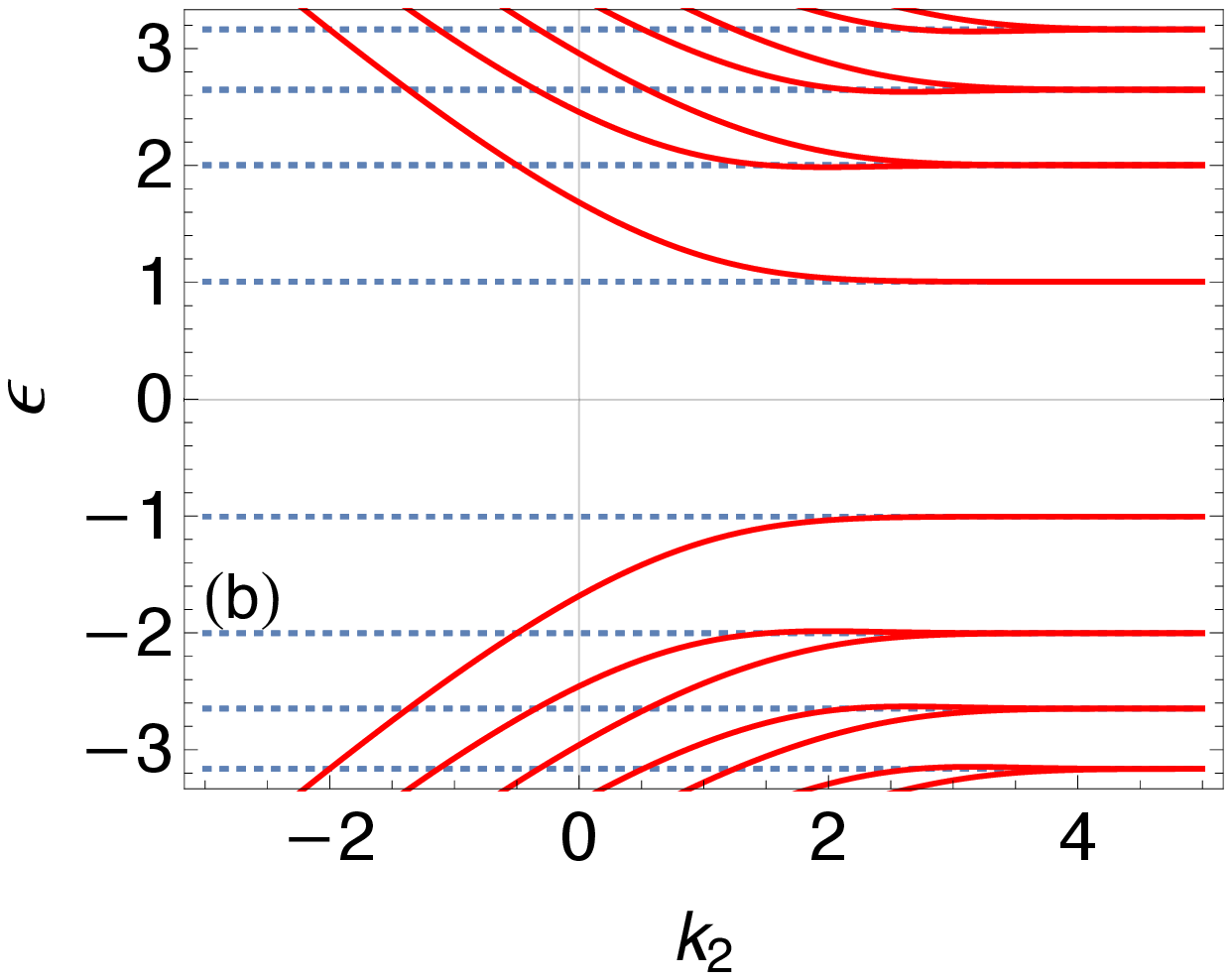}
\end{tabular}
\caption{
Edge (and bulk) states of the BBH Dirac \cin{insulator} with (a) $m_1=1$ and (b) $m_1=-1$. Other parameters used are
$m_2=0.1$ and $B=1.5$. 
The dashed lines are the bulk spectrum of the BBH Dirac \cin{insulator} in a magnetic field,
$\pm m_1$ and 
$\pm\sqrt{(\varepsilon_{0,n}^\pm)^2+m_2^2}$.
}
\label{f:bbh_edge}
\end{center}
\end{figure}

We first mention that considering Eq. (\ref{BBHHam}),
the bulk spectrum is given by $\pm\sqrt{2eBn+m^2}$ ($n=0,1,\cdots$), 
where $m^2=m_1^2+m_2^2$.
Therefore, the bulk gap at zero energy is given by $2m$.

Next, let us consider the system defined in the region $x_1\geq0$.
Let $\psi_{0,\nu}(x_1,k_2)$ be the wave function (\ref{WavFunMag}) of ${\cal H}_0$, on which
 the boundary condition (\ref{BouCon}) [i.e., (\ref{BouEqu})] with $\theta=0$ is imposed.
 Then,
\begin{alignat}1
\psi_{\nu\pm}(x_1,k_2)=
\left(\begin{array}{l}\varepsilon\psi_{0,\nu}(x_1,k_2)\\(\varepsilon_{0,\nu}+im_2)\psi_{0,\nu}(x_1,k_2)\end{array}\right),
\end{alignat}
is the wave function of (\ref{BBHHam}) with energy $\varepsilon=\pm\sqrt{\varepsilon_{0,\nu}^2+m_2^2}$ 
satisfying Eq. (\ref{BBHDirEqu}).

In Fig. \ref{f:bbh_edge}, we show an example of the spectrum of the BBH Dirac \cin{insulator} in a magnetic field.
\cinb{
These figures indeed reproduce the spectra of the BBH model in Fig. \ref{f:bbh_lat_edge}.
}
The case with small $m_2$ is shown in Fig. \ref{f:bbh_edge} (a), in which the gap closing of the edge spectrum,
i.e., the boundary-obstruction at $m_2=0$ is manifest. As discussed in Sec. \ref{s:edge_and_bulk}, the edge state of ${\cal H}_0$
 can be basically given by Eq. (\ref{EffEdgeHam}), implying that an effective Hamiltonian for the edge state 
at $k_2\lesssim0$ of the present BBH Dirac \cin{insulator} is also given by  Eq. (\ref{BBHEdgEff}).
Thus, even in a magnetic field, the gap closing of the edge state induces the \cin{topological} change associated 
with a corner state. This boundary obstruction occurs with keeping the bulk gap $2m$ open. 
On the other hand, in the case of $m_1<0$, no gap closing is observed, as is expected in Fig. \ref{f:bbh_edge} (b).
Therefore, we conclude that \cin{the BBH Dirac insulator} in a magnetic field reproduces the BOTP 
\cinb{ 
transition
} 
of the BBH model,
\cinb{
although the direct calculations of the topological invariants by the use of the Dirac insulator model
in the presence of a magnetic field are impossible.
}

\section{Summary and discussion}\label{s:sum}
\cinb{
To construct an effective theory of the BOTP transition of the BBH model in a magnetic field,
we investigated an effective Dirac fermion model \cin{with two kinds of mass terms} in the continuum limit around $\pi$ flux. 
}
We emphasized the importance of the boundary condition for the Dirac fermion to obtain the 
edge states: We argued the boundary condition from the point of view of
the lattice termination, symmetry, and the Hermiticity condition. 
We first solved the edge states for the conventional 2D \cin{massive} Dirac fermion 
in the absence/presence of  a magnetic field imposing a generic boundary condition.
Using these,
we next derive the edge states of the BBH Dirac \cin{insulator} model.
The gapped edge states of the BBH Dirac \cin{insulator}
show the gap closing \cin{at the transition point from the HOTI phase to the trivial phase}
which is nothing but the BOTP
\cinb{
transition.
}
This occurs even in the presence of a magnetic field, in which the edge states associated with 
the unpaired Landau level causes the BOTP
\cinb{
transition.
}

The result in this paper may be limited within small magnetic fields around $\pi$ flux, since
we use the linear dispersion approximation of the BBH model.
Indeed, as discussed in Sec. \ref{s:bbh_lat}, Fig. \ref{f:BOTP} shows at least in $5\pi/6\leq\phi(\leq7\pi/6)$, the BOTP exists, 
but in $\phi\leq2\pi/3$, the \cin{higher-order topological phase }transition accompanies bulk gap closings. 
It then follows that the description \cin{by the BBH Dirac insulator} 
cannot be extended into such a region. 
It may be an interesting future problem to clarify the nature of this phase.
 
\acknowledgements

This work was supported in part by Grants-in-Aid for Scientific Research Numbers 17K05563 and 17H06138
from the Japan Society for the Promotion of Science.

\appendix

\section{Parabolic cylinder functions $D_\nu(z)$}\label{s:app}

\label{s:D}
The solutions $D_\nu(z)$ of the equation
\begin{alignat}1
\label{webeq}
\left(\frac{d^2}{dz^2}+\nu+\frac{1}{2}-\frac{1}{4}z^2\right)D_\nu(z)=0,
\end{alignat}
are called parabolic cylinder functions \cite{high-tran-2}.
These functions obey
\begin{alignat}1
&\left(\frac{d}{dz}+\frac{z}{2}\right)D_\nu(z)=\nu D_{\nu-1}(z),
\nonumber\\
&\left(\frac{d}{dz}-\frac{z}{2}\right)D_\nu(z)=- D_{\nu+1}(z).
\label{CreAniWav}
\end{alignat}
Other linearly independent solution of Eq. (\ref{webeq}) is $D_\nu(-z)$ if $\nu$ is not an integer, or
$D_{-\nu-1}(iz)$ for any $\nu$. 
When $\nu$ is a non-negative integer, $\nu=n\equiv0,1,\cdots$, $D_{n}(z)$ corresponds to 
the familiar wave function of the 1D harmonic oscillator,
\begin{alignat}1
\label{webher}
&D_{n}(z)=2^{-\frac{1}{2}n}e^{-\frac{1}{4}z^2}H_{n}(2^{-\frac{1}{2}}z),
\end{alignat}
where $H_n(z)$ is the Hermite polynomial of degree $n$.
The asymptotic behavior of $D_\nu(z)$ for large values of $|z|$ and a fixed value of $\nu ~(\ne n)$ is 
\begin{alignat}1
\label{asy1}
D_{\nu}(z)&=z^{\nu}e^{-\frac{1}{4}z^2}\left(1+O(|z|^{-2})\right),
\nonumber\\
&\qquad(|\arg z|<3\pi/4),
\end{alignat}
whereas
\begin{alignat}1
\label{asy2}
D_{\nu}(z)&=z^{\nu}e^{-\frac{1}{4}z^2}\left(1+O(|z|^{-2})\right)
\nonumber\\ 
&-\frac{(2\pi)^\frac{1}{2}}{\Gamma(-\nu)}e^{i\nu\pi}z^{-\nu-1}e^{\frac{1}{4}z^2 }\left(1+O(|z|^{-2})\right),
\nonumber\\ 
&\qquad(\pi/4<\arg z<5\pi/4).
\end{alignat}
Therefore, Eq. (\ref{asy2}) tells that $D_\nu(z)$ diverges as $z^{-\nu-1}e^{\frac{1}{4}z^2}$ 
for real negative $z$, $z\rightarrow-\infty$, if $\nu~(\ne n)$.


\begin{thebibliography}{29}
\expandafter\ifx\csname natexlab\endcsname\relax\def\natexlab#1{#1}\fi
\expandafter\ifx\csname bibnamefont\endcsname\relax
  \def\bibnamefont#1{#1}\fi
\expandafter\ifx\csname bibfnamefont\endcsname\relax
  \def\bibfnamefont#1{#1}\fi
\expandafter\ifx\csname citenamefont\endcsname\relax
  \def\citenamefont#1{#1}\fi
\expandafter\ifx\csname url\endcsname\relax
  \def\url#1{\texttt{#1}}\fi
\expandafter\ifx\csname urlprefix\endcsname\relax\def\urlprefix{URL }\fi
\providecommand{\bibinfo}[2]{#2}
\providecommand{\eprint}[2][]{\url{#2}}

\bibitem[{\citenamefont{Benalcazar
  et~al.}(2017{\natexlab{a}})\citenamefont{Benalcazar, Bernevig, and
  Hughes}}]{Benalcazar:2017aa}
\bibinfo{author}{\bibfnamefont{W.~A.} \bibnamefont{Benalcazar}},
  \bibinfo{author}{\bibfnamefont{B.~A.} \bibnamefont{Bernevig}},
  \bibnamefont{and} \bibinfo{author}{\bibfnamefont{T.~L.}
  \bibnamefont{Hughes}}, \bibinfo{journal}{Physical Review B}
  \textbf{\bibinfo{volume}{96}}, \bibinfo{pages}{245115}
  (\bibinfo{year}{2017}{\natexlab{a}}).

\bibitem[{\citenamefont{Benalcazar
  et~al.}(2017{\natexlab{b}})\citenamefont{Benalcazar, Bernevig, and
  Hughes}}]{Benalcazar:2017ab}
\bibinfo{author}{\bibfnamefont{W.~A.} \bibnamefont{Benalcazar}},
  \bibinfo{author}{\bibfnamefont{B.~A.} \bibnamefont{Bernevig}},
  \bibnamefont{and} \bibinfo{author}{\bibfnamefont{T.~L.}
  \bibnamefont{Hughes}}, \bibinfo{journal}{Science}
  \textbf{\bibinfo{volume}{357}}, \bibinfo{pages}{61}
  (\bibinfo{year}{2017}{\natexlab{b}}).

\bibitem[{\citenamefont{Schindler et~al.}(2018)\citenamefont{Schindler, Cook,
  Vergniory, Wang, Parkin, Bernevig, and Neupert}}]{Schindler:2018ab}
\bibinfo{author}{\bibfnamefont{F.}~\bibnamefont{Schindler}},
  \bibinfo{author}{\bibfnamefont{A.~M.} \bibnamefont{Cook}},
  \bibinfo{author}{\bibfnamefont{M.~G.} \bibnamefont{Vergniory}},
  \bibinfo{author}{\bibfnamefont{Z.}~\bibnamefont{Wang}},
  \bibinfo{author}{\bibfnamefont{S.~S.~P.} \bibnamefont{Parkin}},
  \bibinfo{author}{\bibfnamefont{B.~A.} \bibnamefont{Bernevig}},
  \bibnamefont{and} \bibinfo{author}{\bibfnamefont{T.}~\bibnamefont{Neupert}},
  \bibinfo{journal}{Science Advances} \textbf{\bibinfo{volume}{4}},
  \bibinfo{pages}{eaat0346} (\bibinfo{year}{2018}).

\bibitem[{\citenamefont{Hayashi}(2018)}]{Hayashi:2018aa}
\bibinfo{author}{\bibfnamefont{S.}~\bibnamefont{Hayashi}},
  \bibinfo{journal}{Communications in Mathematical Physics}
  \textbf{\bibinfo{volume}{364}}, \bibinfo{pages}{343} (\bibinfo{year}{2018}).

\bibitem[{\citenamefont{Hashimoto and Kimura}(2016)}]{Hashimoto:2016aa}
\bibinfo{author}{\bibfnamefont{K.}~\bibnamefont{Hashimoto}} \bibnamefont{and}
  \bibinfo{author}{\bibfnamefont{T.}~\bibnamefont{Kimura}},
  \bibinfo{journal}{Physical Review B} \textbf{\bibinfo{volume}{93}},
  \bibinfo{pages}{195166} (\bibinfo{year}{2016}).

\bibitem[{\citenamefont{Hashimoto et~al.}(2017)\citenamefont{Hashimoto, Wu, and
  Kimura}}]{Hashimoto:2017aa}
\bibinfo{author}{\bibfnamefont{K.}~\bibnamefont{Hashimoto}},
  \bibinfo{author}{\bibfnamefont{X.}~\bibnamefont{Wu}}, \bibnamefont{and}
  \bibinfo{author}{\bibfnamefont{T.}~\bibnamefont{Kimura}},
  \bibinfo{journal}{Physical Review B} \textbf{\bibinfo{volume}{95}},
  \bibinfo{pages}{165443} (\bibinfo{year}{2017}).

\bibitem[{\citenamefont{Langbehn et~al.}(2017)\citenamefont{Langbehn, Peng,
  Trifunovic, von Oppen, and Brouwer}}]{Langbehn:2017aa}
\bibinfo{author}{\bibfnamefont{J.}~\bibnamefont{Langbehn}},
  \bibinfo{author}{\bibfnamefont{Y.}~\bibnamefont{Peng}},
  \bibinfo{author}{\bibfnamefont{L.}~\bibnamefont{Trifunovic}},
  \bibinfo{author}{\bibfnamefont{F.}~\bibnamefont{von Oppen}},
  \bibnamefont{and} \bibinfo{author}{\bibfnamefont{P.~W.}
  \bibnamefont{Brouwer}}, \bibinfo{journal}{Physical Review Letters}
  \textbf{\bibinfo{volume}{119}}, \bibinfo{pages}{246401}
  (\bibinfo{year}{2017}).

\bibitem[{\citenamefont{Song et~al.}(2017)\citenamefont{Song, Fang, and
  Fang}}]{Song:2017aa}
\bibinfo{author}{\bibfnamefont{Z.}~\bibnamefont{Song}},
  \bibinfo{author}{\bibfnamefont{Z.}~\bibnamefont{Fang}}, \bibnamefont{and}
  \bibinfo{author}{\bibfnamefont{C.}~\bibnamefont{Fang}},
  \bibinfo{journal}{Physical Review Letters} \textbf{\bibinfo{volume}{119}},
  \bibinfo{pages}{246402} (\bibinfo{year}{2017}).

\bibitem[{\citenamefont{Ezawa}(2018{\natexlab{a}})}]{Ezawa:2018aa}
\bibinfo{author}{\bibfnamefont{M.}~\bibnamefont{Ezawa}},
  \bibinfo{journal}{Physical Review Letters} \textbf{\bibinfo{volume}{120}},
  \bibinfo{pages}{026801} (\bibinfo{year}{2018}{\natexlab{a}}).

\bibitem[{\citenamefont{Ezawa}(2018{\natexlab{b}})}]{Ezawa:2018ab}
\bibinfo{author}{\bibfnamefont{M.}~\bibnamefont{Ezawa}},
  \bibinfo{journal}{Physical Review B} \textbf{\bibinfo{volume}{98}},
  \bibinfo{pages}{045125} (\bibinfo{year}{2018}{\natexlab{b}}).

\bibitem[{\citenamefont{Liu and Wakabayashi}(2017)}]{Liu:2017aa}
\bibinfo{author}{\bibfnamefont{F.}~\bibnamefont{Liu}} \bibnamefont{and}
  \bibinfo{author}{\bibfnamefont{K.}~\bibnamefont{Wakabayashi}},
  \bibinfo{journal}{Physical Review Letters} \textbf{\bibinfo{volume}{118}},
  \bibinfo{pages}{076803} (\bibinfo{year}{2017}).

\bibitem[{\citenamefont{Khalaf}(2018)}]{Khalaf:2018cr}
\bibinfo{author}{\bibfnamefont{E.}~\bibnamefont{Khalaf}},
  \bibinfo{journal}{Physical Review B} \textbf{\bibinfo{volume}{97}},
  \bibinfo{pages}{205136} (\bibinfo{year}{2018}).

\bibitem[{\citenamefont{Matsugatani and Watanabe}(2018)}]{Matsugatani:2018aa}
\bibinfo{author}{\bibfnamefont{A.}~\bibnamefont{Matsugatani}} \bibnamefont{and}
  \bibinfo{author}{\bibfnamefont{H.}~\bibnamefont{Watanabe}},
  \bibinfo{journal}{Physical Review B} \textbf{\bibinfo{volume}{98}},
  \bibinfo{pages}{205129} (\bibinfo{year}{2018}).

\bibitem[{\citenamefont{Fukui and Hatsugai}(2018)}]{Fukui:2018aa}
\bibinfo{author}{\bibfnamefont{T.}~\bibnamefont{Fukui}} \bibnamefont{and}
  \bibinfo{author}{\bibfnamefont{Y.}~\bibnamefont{Hatsugai}},
  \bibinfo{journal}{Physical Review B} \textbf{\bibinfo{volume}{98}},
  \bibinfo{pages}{035147} (\bibinfo{year}{2018}).

\bibitem[{\citenamefont{C{\u a}lug{\u a}ru et~al.}(2019)\citenamefont{C{\u
  a}lug{\u a}ru, Juri{\v c}i{\'c}, and Roy}}]{Calugaru:2019aa}
\bibinfo{author}{\bibfnamefont{D.}~\bibnamefont{C{\u a}lug{\u a}ru}},
  \bibinfo{author}{\bibfnamefont{V.}~\bibnamefont{Juri{\v c}i{\'c}}},
  \bibnamefont{and} \bibinfo{author}{\bibfnamefont{B.}~\bibnamefont{Roy}},
  \bibinfo{journal}{Physical Review B} \textbf{\bibinfo{volume}{99}},
  \bibinfo{pages}{041301} (\bibinfo{year}{2019}).

\bibitem[{\citenamefont{Khalaf et~al.}(2019)\citenamefont{Khalaf, Benalcazar,
  Hughes, and Queiroz}}]{1908.00011}
\bibinfo{author}{\bibfnamefont{E.}~\bibnamefont{Khalaf}},
  \bibinfo{author}{\bibfnamefont{W.~A.} \bibnamefont{Benalcazar}},
  \bibinfo{author}{\bibfnamefont{T.~L.} \bibnamefont{Hughes}},
  \bibnamefont{and} \bibinfo{author}{\bibfnamefont{R.}~\bibnamefont{Queiroz}}
  (\bibinfo{year}{2019}), \eprint{arXiv:1908.00011}.

\bibitem[{\citenamefont{Su et~al.}(1979)\citenamefont{Su, Schrieffer, and
  Heeger}}]{Su:1979aa}
\bibinfo{author}{\bibfnamefont{W.~P.} \bibnamefont{Su}},
  \bibinfo{author}{\bibfnamefont{J.~R.} \bibnamefont{Schrieffer}},
  \bibnamefont{and} \bibinfo{author}{\bibfnamefont{A.~J.}
  \bibnamefont{Heeger}}, \bibinfo{journal}{Physical Review Letters}
  \textbf{\bibinfo{volume}{42}}, \bibinfo{pages}{1698} (\bibinfo{year}{1979}).

\bibitem[{\citenamefont{Wheeler et~al.}(2019)\citenamefont{Wheeler, Wagner, and
  Hughes}}]{Wheeler:2019tg}
\bibinfo{author}{\bibfnamefont{W.~A.} \bibnamefont{Wheeler}},
  \bibinfo{author}{\bibfnamefont{L.~K.} \bibnamefont{Wagner}},
  \bibnamefont{and} \bibinfo{author}{\bibfnamefont{T.~L.}
  \bibnamefont{Hughes}}, \bibinfo{journal}{Physical Review B}
  \textbf{\bibinfo{volume}{100}}, \bibinfo{pages}{245135}
  (\bibinfo{year}{2019}).

\bibitem[{\citenamefont{Otaki and Fukui}(2019)}]{Otaki:2019aa}
\bibinfo{author}{\bibfnamefont{Y.}~\bibnamefont{Otaki}} \bibnamefont{and}
  \bibinfo{author}{\bibfnamefont{T.}~\bibnamefont{Fukui}},
  \bibinfo{journal}{Physical Review B} \textbf{\bibinfo{volume}{100}},
  \bibinfo{pages}{245108} (\bibinfo{year}{2019}).

\bibitem[{\citenamefont{Fukui}(2019)}]{Fukui:2019aa}
\bibinfo{author}{\bibfnamefont{T.}~\bibnamefont{Fukui}},
  \bibinfo{journal}{Physical Review B} \textbf{\bibinfo{volume}{99}},
  \bibinfo{pages}{165129} (\bibinfo{year}{2019}).

\bibitem[{\citenamefont{Jackiw and Rossi}(1981)}]{JackiwRossi:1981}
\bibinfo{author}{\bibfnamefont{R.}~\bibnamefont{Jackiw}} \bibnamefont{and}
  \bibinfo{author}{\bibfnamefont{P.}~\bibnamefont{Rossi}},
  \bibinfo{journal}{Nucl. Phys. B} \textbf{\bibinfo{volume}{190}},
  \bibinfo{pages}{681} (\bibinfo{year}{1981}).

\bibitem[{\citenamefont{Witten}(2015)}]{1510.07698}
\bibinfo{author}{\bibfnamefont{E.}~\bibnamefont{Witten}}
  (\bibinfo{year}{2015}), \eprint{arXiv:1510.07698}.

\bibitem[{\citenamefont{Ishikawa}(1985)}]{Ishikawa:1985uq}
\bibinfo{author}{\bibfnamefont{K.}~\bibnamefont{Ishikawa}},
  \bibinfo{journal}{Physical Review D} \textbf{\bibinfo{volume}{31}},
  \bibinfo{pages}{1432} (\bibinfo{year}{1985}).

\bibitem[{\citenamefont{L.~Alvarez-Gaum\'e}(1984)}]{alvarez85}
\bibinfo{author}{\bibfnamefont{} \bibnamefont{L.~Alvarez-Gaum\'e},
  \bibfnamefont{S.~Della Pietra, and G. Moore}}, \bibinfo{journal}{Annals of Physics}
  \textbf{\bibinfo{volume}{163}}, \bibinfo{pages}{288} (\bibinfo{year}{1984}).

\bibitem[{\citenamefont{Ishikawa}(1984)}]{Ishikawa:1984aa}
\bibinfo{author}{\bibfnamefont{K.}~\bibnamefont{Ishikawa}},
  \bibinfo{journal}{Physical Review Letters} \textbf{\bibinfo{volume}{53}},
  \bibinfo{pages}{1615} (\bibinfo{year}{1984}).

\bibitem[{\citenamefont{Semenoff}(1984)}]{Semenoff:1984aa}
\bibinfo{author}{\bibfnamefont{G.~W.} \bibnamefont{Semenoff}},
  \bibinfo{journal}{Physical Review Letters} \textbf{\bibinfo{volume}{53}},
  \bibinfo{pages}{2449} (\bibinfo{year}{1984}).

\bibitem[{\citenamefont{Redlich}(1984{\natexlab{a}})}]{Redlich:1984kx}
\bibinfo{author}{\bibfnamefont{A.~N.} \bibnamefont{Redlich}},
  \bibinfo{journal}{Physical Review Letters} \textbf{\bibinfo{volume}{52}},
  \bibinfo{pages}{18 } (\bibinfo{year}{1984}{\natexlab{a}}).

\bibitem[{\citenamefont{Redlich}(1984{\natexlab{b}})}]{Redlich:1984uq}
\bibinfo{author}{\bibfnamefont{A.~N.} \bibnamefont{Redlich}},
  \bibinfo{journal}{Physical Review D} \textbf{\bibinfo{volume}{29}},
  \bibinfo{pages}{2366 } (\bibinfo{year}{1984}{\natexlab{b}}).

\bibitem[{\citenamefont{Bateman}(1953)}]{high-tran-2}
\bibinfo{author}{\bibfnamefont{H.}~\bibnamefont{Bateman}},
  \emph{\bibinfo{title}{Higher transcendental functions}},
  vol.~\bibinfo{volume}{II} (\bibinfo{publisher}{McGraw-Hill Book Company},
  \bibinfo{year}{1953}).

\end{thebibliography}

\end{document}